\documentclass[epj,final,nopacs]{myownsvjour}
\usepackage{graphicx}
\usepackage{bm}
\newcommand{\beq}{\begin{equation}}
\newcommand{\beqa}{\begin{eqnarray}}
\newcommand{\eeq}{\end{equation}}
\newcommand{\eeqa}{\end{eqnarray}}

\renewcommand{\a}{\alpha}
\renewcommand{\b}{\beta}
\newcommand{\abs}[1]{\vert#1\vert}
\newcommand{\ave}[1]{{\overline{#1}}}
\newcommand{\avg}{{\rm ave}}
\newcommand{\e}{{\rm e}}
\newcommand{\eps}{\varepsilon}
\newcommand{\frad}[2]{\displaystyle{\displaystyle#1\over\displaystyle#2}}
\renewcommand{\l}{\lambda}
\newcommand{\lra}{\longleftrightarrow}
\newcommand{\m}[1]{{\bm#1}}
\renewcommand{\max}{{\rm max}}
\newcommand{\mean}[1]{\langle#1\rangle}
\renewcommand{\min}{{\rm min}}
\newcommand{\mod}{\mathop{\;\rm mod\;}\nolimits}
\renewcommand{\o}{\omega}
\newcommand{\prob}{\mathop{\rm Prob}\nolimits}
\newcommand{\q}{\kappa}
\renewcommand{\r}{\rho}
\newcommand{\s}{\sigma}
\newcommand{\uni}{{\rm uni}}

\newcommand{\Frac}{\mathop{\rm Frac}\nolimits}
\newcommand{\Int}{\mathop{\rm Int}\nolimits}

\begin{document}

\title{Parrondo games as disordered systems}

\author{Jean-Marc Luck}

\institute{Institut de Physique Th\'eorique,
Universit\'e Paris-Saclay, CEA and CNRS, 91191 Gif-sur-Yvette, France.\\
\email{jean-marc.luck@ipht.fr}}

\date{}

\abstract{
Parrondo's paradox refers to the counter-intuitive situation where a winning strategy
results from a suitable combination of losing ones.
Simple stochastic games exhibiting this paradox have been introduced
around the turn of the millennium.
The common setting of these Parrondo games is that two rules, $A$ and $B$,
are played at discrete time steps,
following either a periodic pattern or an aperiodic one, be it deterministic or random.
These games can be mapped onto 1D random walks.
In capital-dependent games,
the probabilities of moving right or left
depend on the walker's position modulo some integer $K$.
In history-dependent games,
each step is correlated with the $Q$ previous ones.
In both cases the gain identifies with the velocity of the walker's ballistic motion,
which depends non-linearly on model parameters,
allowing for the possibility of Parrondo's paradox.
Calculating the gain involves products of non-commuting Markov matrices,
which are somehow analogous to the transfer matrices
used in the physics of 1D disordered systems.
Elaborating upon this analogy,
we study a paradigmatic Parrondo game of each class
in the neutral situation where each rule, when played alone, is fair.
The main emphasis of this systematic approach
is on the dependence of the gain on the remaining parameters and,
above all, on the game, i.e., the rule pattern,
be it periodic or aperiodic, deterministic or random.
One of the most original sides of this work
is the identification of weak-contrast regimes for capital-dependent
and history-dependent Parrondo games,
and a detailed quantitative investigation of the gain in the latter scaling regimes.
}

\maketitle

\section{Introduction}
\label{intro}

Parrondo's paradox
refers to the counter-intuitive situation where a winning strategy
results from a suitable combination of losing ones.
Simple stochastic games exhibiting this paradox have been introduced
by Parrondo and collaborators around the turn of the millennium~\cite{HA2,MC,HA1,HAT,PHA}.
References~\cite{R1,R2,R3,R4} provide comprehensive reviews
of early developments of Parrondo games,
including historical aspects and extensive discussions of their paradoxical nature.
Parrondo games were originally devised as discrete analogues of Brownian ratchets.
The latter ratchets
are extensions of Feynman's celebrated thermal ratchet~\cite{fey} to the microscopic scale,
aimed at modeling the force-free motion of molecular motors~\cite{AP,mag,AB}.
Flashing Brownian ratchets consist of a point particle
undergoing Brownian diffusion on the line
under the effect of a periodic potential
which is both spatially asymmetric and periodically modulated in time.
The interplay of these two properties breaks detailed balance.
Under generic circumstances,
it yields a rectification of thermal noise
and induces a steady ballistic motion of the particle
(see~\cite{JAP,rei} for reviews).

Parrondo games belong to the realm of Markovian games of chance.
The usual setting is that two stochastic rules, denoted as $A$ and $B$,
are played at discrete time steps in a specific order,
following a periodic pattern such as $ABBABB\dots$
or an aperiodic one, either deterministic or random.
It is advantageous to describe Parrondo games within the framework of a random walker
occupying the sites of an infinite 1D lattice
and moving to neighboring sites at discrete time steps
according to the above stochastic rules.
The discrete position $n_t$ of the walker at integer time $t$
identifies with the capital of the player.
In the generic situation where the walker's motion is ballistic,
its velocity yields the gain $G$ of the player per time step:
\beq
G=\lim_{t\to\infty}\frac{n_t}{t}.
\label{Gdef}
\eeq
Parrondo's paradox holds whenever the chosen game
(rule pattern) yields a positive gain,
whereas each rule, when played alone, either is fair or has a negative~gain:
\beq
\hbox{Parrondo's paradox:}\quad\left\{G>0,\ G_A\le0,\ G_B\le0\right\}.
\label{pdef}
\eeq

There are two main classes of Parrondo games.
The first class is referred to as {\it capital-dependent games}.
The rules, either $A$ or $B$ or both, depend explicitly on the walker's position
(i.e., the player's capital)~$n_t \mod K$,
where $K$ is some fixed integer\footnote{$n \mod K=0,\dots,K-1$ is the rest
of the Euclidean division of $n$ by $K$.}.
The game originally proposed by Parrondo~\cite{HA2,MC,HA1,HAT} corresponds to $K=3$.
Parrondo's paradox also holds for some specific models with $K=2$,
where Rule $B$ depends on the parity of the player's capital~\cite{CV,WXW}.
A second class of Parrondo games, referred to as {\it history-dependent games}~\cite{PHA,KJ,EL},
has also been considered,
even though it has not become as popular as capital-dependent games.
There, the complexity of the dynamics originates in a memory effect between successive steps.
The probability for the walker to move right or left is now independent of its position $n_t$,
but it depends on the $Q$ previous steps,
in a way that is different for Rules~$A$ and $B$.
Parrondo's paradox already holds in some cases for $Q=1$,
and more generally for $Q=2$~\cite{CV}.

Consider for the time being a random walker on an infinite 1D lattice,
with time-dependent probabilities of moving to neighboring sites.
Let $p_t$ (resp.~$q_t=1-p_t$) be the probability that the walker moves to the right
(resp.~to the left) at time $t$.
The mean position of the walker at time~$t$ reads
\beq
\mean{n_t}=n_0+\sum_{s=1}^t(2p_s-1).
\eeq
This expression only depends on the sum of the probability differences $p_s-q_s=2p_s-1$,
and not on the order in which single steps are performed.
In other words, elementary steps commute with each other.
In the case of an annealed disorder,
where the time-dependent probabilities $p_t$ are themselves drawn from some distribution,
the velocity of the walker is self-averaging and reads
\beq
G=2\ave{p}-1.
\eeq
The notations for averages used throughout this paper
follow the usual conventions of the theory of disordered systems.
Brackets, $\mean{\dots}$, denote an average over realizations of the Markov process,
i.e., over histories of the random walker,
whereas a bar, $\ave{\dots\vphantom{m}}$, denotes
an annealed average over the distribution of the probabilities defining the Markov process,
whenever the latter are themselves random.

In the case of Parrondo games,
the existence of internal degrees of freedom
(the walker's position $n \mod K$ for capi\-tal-dependent games,
or the $Q$ previous steps for history-depen\-dent games)
makes the corresponding random walk non-trivial.
The gain $G$, i.e., the walker's velocity,
depends non-linearly on model parameters,
allowing for the possibility of Parrondo's paradox, defined by the inequalities~(\ref{pdef}).
Parrondo games can be viewed as inhomogeneous Markov chains~\cite{R1,R2,R3},
whose study involves products of non-commuting Markov matrices
acting on a finite-dimensional linear space with dimension
\beq
d=K\mbox{ or }d=2^Q,
\eeq
encoding internal degrees of freedom.
These products of Markov matrices are somehow temporal analogues
of the spatial products of non-commuting transfer matrices
that are ubiquitous in investigations of 1D disordered systems
(see~\cite{BL,CPV,JMAlea,pen,CT1,CT2} for reviews).

The goal of the present work is to elaborate on this analogy
and to study Parrondo games by means of various analytical techniques
freely inspired by the theory of~1D disordered systems.
This line of thought allows us to deal with capital-dependent and history-dependent
games on the same footing,
and yields a wealth of new results on both classes of Parrondo games.
We consider capital-dependent games in Sections~\ref{I} and~\ref{Iweak}
and history-depen\-dent games in Sections~\ref{II} and~\ref{IIweak}.
We choose for definiteness to work with one paradigmatic example of each class.
Most of the time,
we focus our attention onto the neutral situation where each rule, when played alone, is fair
($G_A=G_B=0$).
The main emphasis of this systematic approach
is on the dependence of the gain $G$ on the remaining free parameters and,
more importantly, on the game, i.e., the rule pattern,
be it periodic or aperiodic, deterministic or random.
One of the most original sides of this work
is the identification of a weak-contrast scaling regime
and its systematic investigation for both classes of games
(Sections~\ref{Iweak} and~\ref{IIweak}).
Section~\ref{disc} contains a brief overview.

\section{Capital-dependent games}
\label{I}

\subsection{Generalities}
\label{Igen}

\begin{figure}[!ht]
\begin{center}
\includegraphics[angle=0,width=.75\linewidth]{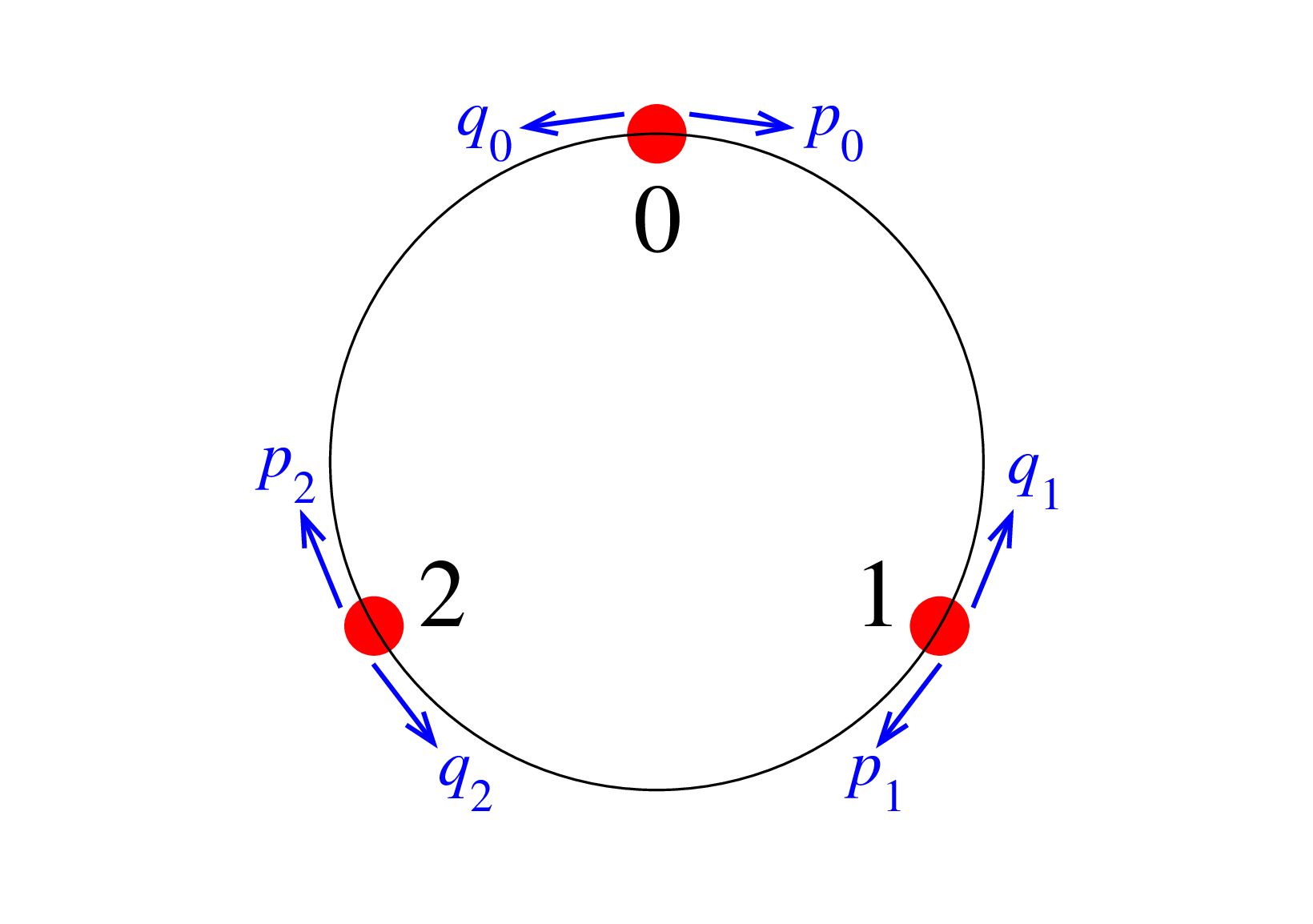}
\caption{
Most general Markovian stochastic rule
of the capital-dependent Parrondo game
in the internal space of the player's capital (i.e., of the walker's position)
$n \mod 3=0$, 1, 2, with the notation $q_n=1-p_n$.}
\label{Icirc}
\end{center}
\end{figure}

The game originally proposed by Parrondo~\cite{HA2,MC,HA1,HAT}
is a prototypical example of a capital-depen\-dent game with $K=3$,
where rules depend on the player's capital (i.e., of the walker's position) mod 3.
It is sufficient to monitor the dynamics of the walker
in the three-dimensional internal space parame\-trized by its position $n \mod 3=0,~1,~2$.
Within this framework, the most general Markovian sto\-chastic rule
is depicted in Figure~\ref{Icirc} and corresponds to the Markov matrix
\beq
\m M=\pmatrix{0&q_1&p_2\cr p_0&0&q_2\cr q_0&p_1&0},
\eeq
with the notation $q_n=1-p_n$.

The standard body of knowledge on Markov chains can be found
in the classical references~\cite{doob,feller,KT,kampen,kelly,stir}.
Hereafter we not pretend at any mathematical rigor.
We shall only need the following general result:
the unique ergodicity of a discrete-time Markov chain,
i.e., essentially the uniqueness of its stationary state,
is ensured by the fact that the corresponding Markov matrix $\m M$
has a simple (i.e., non-degenerate) unit eigenvalue,
while all other eigenvalues are strictly less than unity in modulus.

We introduce the time-dependent state vector
\beq
\m\phi_t=\pmatrix{X_t\cr Y_t\cr Z_t},
\eeq
where $X_t$, $Y_t$ and $Z_t$ are the probabilities that the walker's position $n \mod 3$
at time $t$ is respectively 0, 1 and 2.

Parrondo's historical game consists of a combination of the following rules~\cite{HA2,MC,HA1,HAT}.

\begin{itemize}

\item[$\bullet$] Rule~$A$.
The three probabilities are equal: $p_0=p_1=p_2=p$.
If Rule~$A$ is played at time~$t$, we have
\beq
\m\phi_t=\m M_A\m\phi_{t-1},
\eeq
with
\beq
\m M_A=\pmatrix{0&q&p\cr p&0&q\cr q&p&0}.
\eeq
If Rule~$A$ is played alone,
the walker executes a uniformly biased random walk.
Its stationary state $\m\phi_A$, such that
\beq
\m\phi_A=\m M_A\m\phi_A,
\eeq
is uniform:
\beq
\m\phi_A=\m\phi_\uni=\frac{1}{3}\pmatrix{1\cr 1\cr 1}.
\label{univec}
\eeq
We have
\beq
G_A=\m J_A\cdot\m\phi_A,
\eeq
where the current vector reads
\beq
\m J_A=(2p-1)\pmatrix{1&1&1},
\eeq
and so
\beq
G_A=2p-1.
\label{IGA}
\eeq

\item[$\bullet$] Rule~$B$.
It is defined by setting $p_2=p_1$, keeping $p_0$ and~$p_1$ as free parameters.
If Rule~$B$ is played at time~$t$, we have
\beq
\m\phi_t=\m M_B\m\phi_{t-1},
\eeq
with
\beq
\m M_B=\pmatrix{0&q_1&p_1\cr p_0&0&q_1\cr q_0&p_1&0}.
\eeq
If Rule~$B$ is played alone,
the stationary state of the system is described by the normalized eigenvector $\m\phi_B$
associated with the unit eigenvalue of~$\m M_B$, such that
\beq
\m\phi_B=\m M_B\m\phi_B.
\label{perron}
\eeq
We thus obtain
\beq
\m\phi_B=\pmatrix{X_B\cr Y_B\cr Z_B},
\eeq
with
\beq
X_B=\frac{1-p_1q_1}{D},\ Y_B=\frac{1-q_0p_1}{D},\ Z_B=\frac{1-p_0q_1}{D}
\eeq
and
\beq
D=3-p_0q_1-q_0p_1-p_1q_1.
\eeq
We have
\beq
G_B=\m J_B\cdot\m\phi_B,
\eeq
where the current vector reads
\beq
\m J_B=\pmatrix{p_0-q_0&p_1-q_1&p_1-q_1},
\eeq
and so
\beq
G_B=\frac{3(p_0p_1^2-q_0q_1^2)}{D}.
\label{IGB}
\eeq

\end{itemize}

The Markov matrices $\m M_A$ and $\m M_B$
generically do not commute with each other.
We have indeed
\beq
\left[\m M_A,\m M_B\right]=(p_1-p_0)\pmatrix{2p-1&0&0\cr q&q&p\cr -p&-q&-p}.
\label{abcomm}
\eeq
The commutator vanishes only for $p_1=p_0$,
i.e., when each rule corresponds to a uniformly biased random walk,
so that the dynamics in internal space can be forgotten.

Hereafter the main focus will be on the neutral situation
where each rule, when played alone, is fair ($G_A=G_B=0$).
In this situation, a given game, such as e.g.~the periodic game $ABBABB\dots$,
exhibits Parrondo's paradox whenever the corresponding gain, denoted $G_{ABB}$, is positive
(see~(\ref{pdef})).
The condition that Rule~$A$ is fair reads
\beq
p=\frac{1}{2},
\label{afair}
\eeq
expressing that the corresponding random walk is unbiased, i.e., symmetric.
The condition that Rule~$B$ is fair yields a relation between $p_0$ and $p_1$,
\beq
p_0=\frac{(1-p_1)^2}{1-2p_1(1-p_1)},
\eeq
leaving one free parameter.
It is advantageous to choose the parametrization
\beq
p_0=\frac{1}{2}-\frac{v}{1+v^2},
\quad
p_1=\frac{1}{2}(1+v),
\label{bfair}
\eeq
where the contrast parameter $v$
in the range $-1<v<1$ provides a measure of the difference between both rules.
The expression (\ref{abcomm}) becomes
\beq
\left[\m M_A,\m M_B\right]=\frac{v(3+v^2)}{4(1+v^2)}\pmatrix{0&0&0\cr 1&1&1\cr -1&-1&-1}.
\eeq

We close this section by a discussion of symmetries.

\begin{itemize}

\item[$\bullet$]
Parity, i.e., the change of sign of the walker's position ($n\lra -n$),
corresponds to changing the orientation of the circle shown in Figure~\ref{Icirc}.
It therefore amounts to exchanging the probabilities as $p\lra q$ for Rule~$A$,
and $p_0\lra q_0$, $p_1\lra q_1$ for Rule~$B$.
In the neutral situation, this amounts to changing~$v$ into its opposite ($v\lra-v$).
The gain~$G$ is therefore an odd function of $v$, irrespective of the game.

\smallskip

\item[$\bullet$]
Time reversal
amounts to the sole reversal of the order of letters for a general game of finite duration,
such as
\beq
ABABBABBB \lra BBBABBABA.
\label{abex}
\eeq
The model is indeed simple enough to ensure that each rule is reversible,
i.e., coincides with its own time-reversed, as soon as it is fair.
This is obvious for Rule~$A$.
For Rule~$B$, the expression (\ref{IGB})
shows that the condition for $G_B$ to vanish is
$p_0p_1^2=q_0q_1^2$.
This is nothing but Kolmogorov's criterion for the Markov chain
defining Rule~$B$ to be reversible
(see e.g.~\cite{kelly,stir}).
There is indeed only one non-trivial cycle
(see Figure~\ref{Icirc}),
and so Kolmogorov's criterion amounts to one single equation.
As a consequence of the above,
the gain $G$ is left unchanged under a reversal of the game,
i.e., of the rule pattern, such as (\ref{abex}).

\end{itemize}

\subsection{Random games}
\label{Iran}

The first situation demonstrating Parrondo's paradox is that of an
(infinitely long) random game,
where at each time step Rule~$B$ is chosen with probability $\r$
and Rule~$A$ with the complementary probability $1-\r$.
In the following, we are only interested in the average gain $\ave{G}$ of this random game,
and so it is sufficient to know the average state vector $\ave{\m\phi}$.
The present problem is therefore easier than the investigation of usual 1D disordered systems,
which requires the evaluation of the Lyapunov exponent of a matrix product
(see~\cite{BL,CPV,JMAlea,pen,CT1,CT2} for reviews).
The time-dependent average state vector $\ave{\m\phi}_t$ obeys a recursion of the form
\beq
\ave{\m\phi}_t=\ave{\m M}\,\ave{\m\phi}_{t-1},
\eeq
where the average Markov matrix,
\beq
\ave{\m M}=(1-\r)\m M_A+\r\m M_B,
\label{avem}
\eeq
has the same functional form as $\m M_B$, albeit with effective parameters~\cite{R2,R3}
\beq
\ave p_0=(1-\r)p+\r p_0,\quad
\ave p_1=(1-\r)p+\r p_1.
\eeq
The average gain $\ave G$ of the random game is obtained by
replacing in (\ref{IGB}) $p_0$ and $p_1$ by the above effective values.

For the uniformly random game ($\r=1/2$),
where at each time step Rules~$A$ and $B$ are chosen with equal probabilities, we obtain
\beq
\ave G=\frac{3((p+p_0)(p+p_1)^2-(q+q_0)(q+q_1)^2)}{D},
\label{Igur}
\eeq
with
\beqa
D=2(12&-&(p+p_0)(q+q_1)-(q+q_0)(p+p_1)
\nonumber\\
&-&(p+p_1)(q+q_1)).
\eeqa
The expression (\ref{Igur}) allows one to measure how rare is Parrondo's paradox.
In the present setting,
it is natural to define the probability of observing Parrondo's paradox
as the volume of the three-dimensional domain in $(p,p_0,p_1)$ space such that
the inequalities (\ref{pdef}) hold, with $G$ given by~(\ref{Igur}).
A numerical integration yields
\beq
\prob(\hbox{Parrondo's paradox})\approx0.000306.
\label{pc}
\eeq
This very small number is in perfect agreement with an earlier estimate~\cite{CNT}.

From now on, until the end of Section~\ref{Iweak},
we restrict the analysis to capital-dependent Parrondo games in the neutral situation
where both rules, when played alone, are fair ($G_A=G_B=0$),
Using the parametrization (\ref{afair}), (\ref{bfair}),
we obtain the following expression for the average gain:
\beq
\ave{G}=\frac{6\r(1-\r^2)v^3}{9(1+v^2)+\r^2v^2(v^2-3)}.
\label{Iaveg}
\eeq

The above result exhibits several features of interest.
It is an odd function of the contrast parameter $v$,
as expected from the above considerations on parity.
The average gain has the sign of $v$, irrespective of $\r$.
Parrondo's paradox therefore holds for all $v>0$
and all non-trivial probabilities ($0<\r<1$).
There is no discrepancy with the tininess of the probability (\ref{pc}),
since we have fixed two of the three model parameters
by focussing our attention onto the neutral situation.
The average gain vanishes as $\r\to0$ and $\r\to1$,
where random games respectively degenerate to Rule~$A$ and Rule~$B$.
It reaches its absolute maximum,
\beq
\ave{G}^\max=8\sqrt{2}-5\sqrt{5}=0.133368,
\eeq
for
\beq
\r=2\sqrt{2}-\sqrt{5}=0.592359
\label{rmaxI1}
\eeq
and $v\to1$.
The latter limit is however singular, as it corresponds to $p_0\to0$ and $p_1\to1$.
In this limit, the Markov matrix $\m M_B$ looses the property of unique ergodicity,
as its eigenvalues become 0 and $\pm1$.

In the weak-contrast regime ($v\to0$), the average gain vanishes cubically.
We shall see in Section~\ref{Iweak} that this cubic law holds for arbitrary games.
We are thus led to introduce the gain amplitude
(or amplitude, for short)
\beq
g=\lim_{v\to0}\frac{G}{v^3}.
\label{gdef}
\eeq
For random games, the expression (\ref{Iaveg}) yields
\beq
\ave{g}=\frac{2\r(1-\r^2)}{3}.
\label{gres}
\eeq
For the uniformly random game ($\r=1/2$),
the average amplitude reads
\beq
\ave{g}=\frac{1}{4}.
\label{gavg}
\eeq
When the probability $\r$ of choosing Rule~$B$ varies between~0 and 1,
the amplitude (\ref{gres}) reaches its maximum
\beq
\ave{g}=\frac{4\sqrt{3}}{27}=0.256600
\label{gavmax}
\eeq
for
\beq
\r=\frac{\sqrt{3}}{3}=0.577350.
\label{rmaxI0}
\eeq

\subsection{Periodic games}
\label{Iper}

In this section we consider periodic games,
i.e., periodic rule patterns,
defined by the infinite repetition of a unit cell $W$ of length $P$,
like e.g.~$W=ABB$, which has period $P=3$.
We shall alternatively consider $W$ as a word consisting of~$P$ letters, $A$ or $B$,
and introduce the symbols
\beq
\left\{\matrix{\tau_n=A\cr \s_n=0\hfill}\right.
\quad\hbox{or}\quad
\left\{\matrix{\tau_n=B\cr \s_n=1,\hfill}\right.
\label{symbs}
\eeq
according to whether the $n$th letter in $W$ is $A$ or $B$.
The stationary state of the game has the same period~$P$ as the game itself.
It is encoded in $P$ state vectors $\m\phi_n$ obeying
\beq
\m\phi_n=\m M_{\tau_n} \m\phi_{n-1}\quad(n=1,\dots,P),
\label{eqper}
\eeq
with periodic boundary conditions ($\m\phi_P=\m\phi_0$).
The associated gain reads
\beq
G_W=\frac{1}{P}\sum_{n=1}^P\m J_{\tau_n}\cdot\m\phi_{n-1},
\eeq
where the current vectors $\m J_A$ and $\m J_B$ are evaluated in the neutral situation,
with parameters~(\ref{afair}),~(\ref{bfair}), i.e.,
\beq
\m J_A=\m 0,\quad
\m J_B=\pmatrix{-\frad{2v}{1+v^2}&v&v}.
\eeq

The recursion (\ref{eqper}) amounts to a system of $3P$ linear equations,
whose solution may be obtained by means of a computer algebra system such as MACSYMA.
The complexity of the expressions of the gain $G$ however grows very rapidly with the period $P$.
We recall that the gain is invariant under cyclic permutations and reversal of the unit cell.
Its expressions for all games with periods 2 and~3 are given below.

\begin{itemize}

\item[$\bullet$] $P=2$.
There is only one non-trivial unit cell with period~2, namely $W=AB$.
The corresponding gain vanishes~\cite{EL}:
\beq
G_{AB}=0.
\eeq
This result comes as a surprise, as it is not dictated by any obvious symmetry.

\item[$\bullet$] $P=3$.
There are two inequivalent unit cells with period 3.
The corresponding gains read
\beqa
G_{AAB}&=&\frac{16v^3}{81+78v^2+v^4},
\\
G_{ABB}&=&\frac{8v^3(7+10v^2-v^4)}{81+204v^2+118v^4-20v^6+v^8}.
\eeqa

\end{itemize}

The above expressions demonstrate that the gain vanishes cubically
in the weak-contrast regime ($v\to0$),
which will be the subject of Section~\ref{Iweak}.
The corresponding amplitudes
$g_{AB}=0$, $g_{AAB}=16/81$ and $g_{ABB}=56/81$
(see~(\ref{gdef}))
are listed in the first three lines of Table~\ref{pg}.

\section{Weak-contrast scaling regime of capital-dependent games}
\label{Iweak}

\subsection{Generalities}
\label{Iwgen}

In the weak-contrast scaling regime ($v\to0$),
both rules are close to symmetric random walks,
so that state vectors are expected to become close to the uniform one,
given by~(\ref{univec}).
It can indeed be checked, in full generality,
that the differences between $Y_n$ or $Z_n$ and $1/3$ are of order~$v$,
whereas the difference between $X_n$ and $1/3$ is of order $v^2$,
and the resulting gain is of order~$v^3$.

Hereafter we use the shorthand notation
\beq
\q=-\frac{1}{2}.
\eeq
Let us focus for a while our attention onto periodic games,
considered in Section~\ref{Iper}.
The matrix recursion (\ref{eqper}) between state vectors $\m\phi_n$
boils down to two coupled linear recursions for the rescaled co-ordinates
\beqa
y_n&=&\lim_{v\to0}\frac{3(Y_n-Z_n)}{v},
\\
x_n&=&\lim_{v\to0}\frac{1-3X_n}{v^2},
\eeqa
namely
\beqa
y_n&=&\q\left(y_{n-1}+6\s_n\right),
\label{Iyr}
\\
x_n&=&\q\left(x_{n-1}-\s_ny_{n-1}\right),
\label{Ixr}
\eeqa
with periodic boundary conditions ($y_P=y_0$, $x_P=x_0$).
The gain amplitude (see~(\ref{gdef})) reads
\beq
g_W
=\lim_{v\to0}\frac{G_W}{v^3}
=\frac{1}{3P}\sum_{n=1}^P\s_n(3x_{n-1}+2).
\label{gsum}
\eeq

The recursions~(\ref{Iyr}),~(\ref{Ixr})
are instrumental in the investigation of the weak-contrast regime.
Their key property is the occurrence of the uniform damping factor $\q$,
whereas the rule pattern, encoded in the symbol $\s_n=0$ or 1,
according to (\ref{symbs}), enters linearly.
The above formalism extends to aperiodic games,
either deterministic or random (see Section~\ref{Iwape}).

\subsection{Random games}
\label{Iwran}

As a first application of the above formalism,
let us revisit random games, already considered in Section~\ref{Iran}.
In~(\ref{Ixr}),~$\s_n$ and $y_{n-1}$ are statistically independent,
and we have $\ave{\s_n}=\r$.
The stationary averages $\ave{y}$ and $\ave{x}$ therefore obey
\beqa
\ave{y}&=&\q\left(\ave{y}+6\r\right),
\\
\ave{x}&=&\q\left(\ave{x}-\r\ave{y}\right),
\eeqa
hence
\beq
\ave{y}=-2\r,
\quad
\ave{x}=-\frac{2\r^2}{3},
\eeq
and
\beq
\ave{g}=\frac{\r}{3}\left(3\ave{x}+2\right)=\frac{2\r(1-\r^2)}{3}.
\eeq
The result (\ref{gres}) is thus recovered.

\subsection{Periodic games}
\label{Iwp}

We now turn to the case of periodic games,
already considered in Section~\ref{Iper}.
The explicit solution to (\ref{Iyr}), (\ref{Ixr}) with periodic boundary conditions reads
\beqa
y_n&=&-\frac{3}{1-\q^P}\sum_{m=0}^{P-1}\q^m\s_{n-m},
\\
x_n&=&-\frac{3}{2(1-\q^P)^2}
\nonumber\\
&\times&\sum_{l,m=0}^{P-1}\q^{l+m}\s_{n-m}\s_{n-l-m-1}.
\eeqa
Inserting the latter expression for $x_n$ into (\ref{gsum}),
we obtain after some algebra
\beqa
g_W&=&\frac{6}{P(1-\q^P)^2}
\nonumber\\
&\times&\sum_{k,l,m=1}^P\q^{l+m}\s_k(1-\s_{k+l}\s_{k-m}).
\label{gfullres}
\eeqa
In the above, all indices of $\s$ symbols are to be understood mod $P$.

The result~(\ref{gfullres}) provides an explicit expression of the gain
of Parrondo's historical game
for an arbitrary periodic rule pattern in the weak-contrast regime.
The cyclic and reversal invariance of the gain appear manifestly.
The extension of the above result to aperiodic games
will be considered in Section~\ref{Iwape}.

For the time being we keep the focus onto periodic games.
For a given period $P$,
(\ref{gfullres}) shows that all amplitudes are rational numbers
whose denominator divides $P(2^P-(-1)^P)^2$.
In the case where the unit cell~$W$ consists of only two blocks,
\beq
W=A^MB^N,
\label{wab}
\eeq
with arbitrary integers $M$, $N\ge1$,
so that $P=M+N$,
the expression (\ref{gfullres}) simplifies to
\beqa
g_{A^MB^N}&=&\frac{4(1-\q^M)}{9P(1-\q^P)^2}
\label{gwab}
\\
&\times&\left(3N\q^N(1-\q^M)+2(1-\q^N)(1-\q^P)\right).
\nonumber
\eeqa
When both block lengths $M$ and $N$ become large,
the amplitude falls off as
\beq
g_{A^MB^N}\approx\frac{8}{9P},
\label{gwasy}
\eeq
up to exponentially small corrections.
This decay law in $1/P$ can be interpreted as follows.
Both rules $A$ and $B$ are fair,
and so only the interfaces between blocks yield some gain.
More generally, when one of the block lengths gets large,
the other one being kept finite, (\ref{gwab}) yields
\beq
g_{A^MB^N}\approx\frac{a_M}{N},\quad
a_M=\frac{8}{9}(1-\q^M)
\label{gaasy}
\eeq
for $N\to\infty$ at fixed $M$, and
\beq
g_{A^MB^N}\approx\frac{b_N}{M},\quad
b_N=\frac{4}{9}(2+(3N-2)\q^N)
\label{gbasy}
\eeq
for $M\to\infty$ at fixed $N$.
Both sequences $a_M$ and $b_N$ converge to $8/9$,
consistently with (\ref{gwasy}),
with exponentially damped oscillations.
The smallest of them are $a_2=2/3$ and $b_3=1/2$,
whereas the largest read $a_1=b_2=4/3$.

We now turn to general features of interest
exhibited by the gain amplitudes of periodic games.
The dependence of $g_W$ on the unit cell $W$
defining the periodic game appears to be very intricate in general.
The result (\ref{gwab}) indeed virtually exhausts all cases where (\ref{gfullres})
yields manageable closed-form expression.

Table~\ref{pg} gives
the exact rational and numerical expressions of the gain amplitude $g_W$
for all periodic games with primitive\footnote{The primitive period $P$ of a periodic sequence
is its smallest positive period.} period $P\le6$.
The explicit result (\ref{gwab}) yields~15 of the 20 expressions given there,
whereas the remaining five cases need a specific evaluation of the triple sum
entering (\ref{gfullres}).
The last column gives the corresponding rotation number $\o$
of the cut-and-project sequence (see Section~\ref{caps}), when applicable.

\begin{table}[!ht]
\begin{center}
\begin{tabular}{|c|r|r|c|}
\hline
$\hfill P \hfill$ & $\hfill W \hfill$ & $\hfill g_W \hfill$ & $\o$ \\
\hline
2& $AB$ & $\hfill 0 \hfill$ & $1/2$ \\
\hline
3& $AAB$ & $16/81=0.197530$ & $1/3$ \\
 & $ABB$ & $56/81=0.691358$ & $2/3$ \\
\hline
4& $ABBB$ & $2/25=0.080000$ & $3/4$ \\
 & $AAAB$ & $4/25=0.160000$ & $1/4$ \\
 & $AABB$ & $6/25=0.240000$ & - \\
\hline
5& $AABBB$ & $56/605=0.092561$ & - \\
 & $AAAAB$ & $80/605=0.132231$ & $1/5$ \\
 & $AAABB$ & $184/605=0.304132$ & - \\
 & $AABAB$ & $208/605=0.343801$ & $2/5$ \\
 & $ABBBB$ & $232/605=0.383471$ & $4/5$ \\
 & $ABABB$ & $488/605=0.806611$ & $3/5$ \\
\hline
6& $AAABBB$ & $324/3969=0.081632$ & - \\
 & $ABABBB$ & $332/3969=0.083648$ & - \\
 & $AAAAAB$ & $440/3969=0.110859$ & $1/6$ \\
 & $AABBBB$ & $548/3969=0.138070$ & - \\
 & $ABBBBB$ & $604/3969=0.152179$ & $5/6$ \\
 & $AAABAB$ & $712/3969=0.179390$ & - \\
 & $AAAABB$ & $820/3969=0.206601$ & - \\
 & $AABABB$ & $1404/3969=0.353741$ & - \\
\hline
\end{tabular}
\caption{
Exact rational and numerical expressions of the gain amplitude $g_W$
of the capital-dependent Parrondo game
with all periodic rules $W$ with primitive period $P\le6$.
For each period~$P$,
unit cells $W$ are ordered according to increasing gains.
Last column: corresponding rational rotation number $\o$
of the cut-and-project sequence (see Section~\ref{caps}), when applicable.}
\label{pg}
\end{center}
\end{table}

\begin{figure}[!ht]
\begin{center}
\includegraphics[angle=0,width=.8\linewidth]{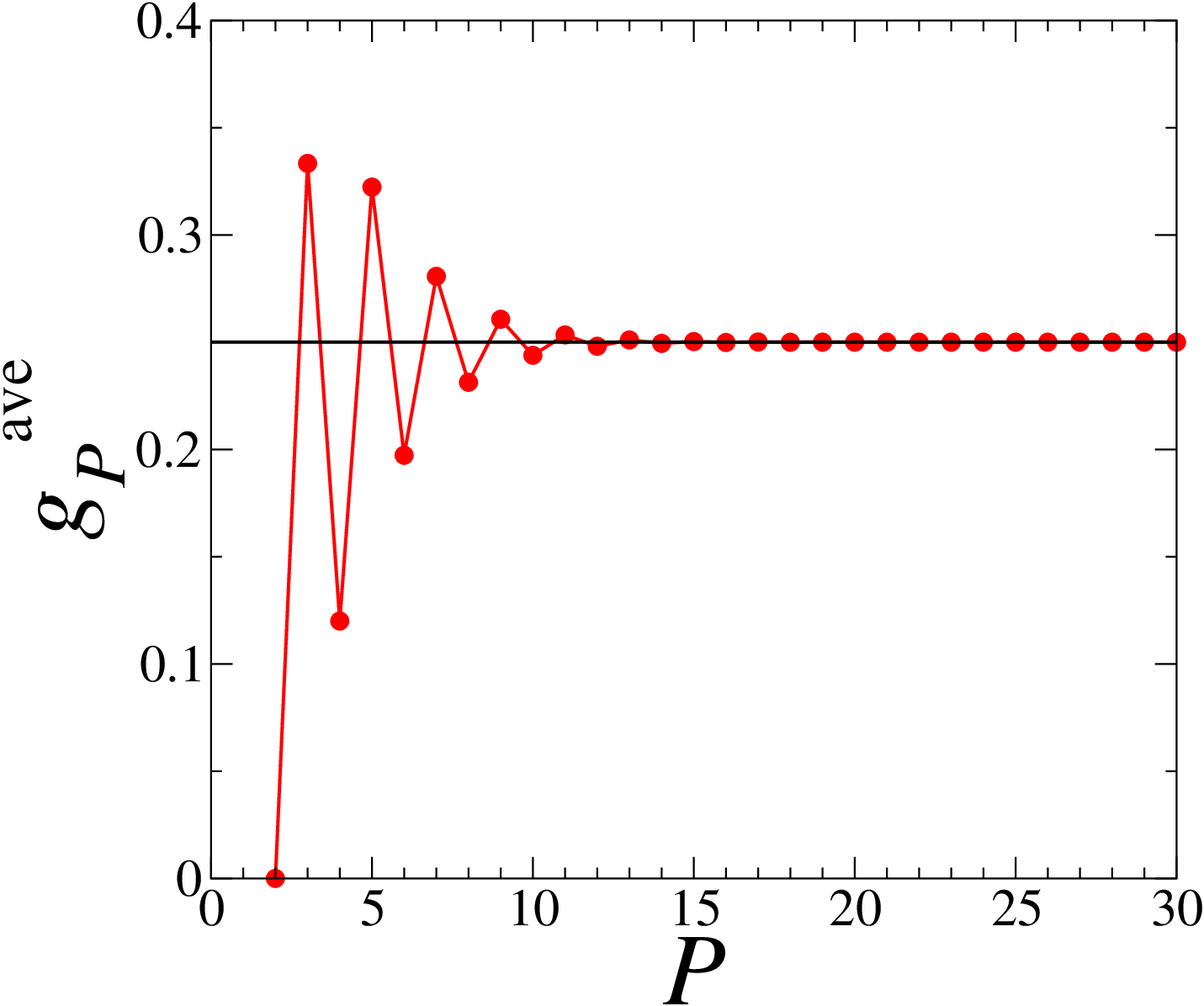}
\caption{
Average gain amplitude $g_P^\avg$ over all periodic
capital-dependent Parrondo games with period $P\le30$.
Horizontal black line: asymptotic limit $1/4$ (see~(\ref{gavg})).}
\label{Iave}
\end{center}
\end{figure}

For a given -- not necessarily primitive -- period $P$,
the~$2^P$ possible unit cells $W$ of length~$P$
can be enumerated by means of a computer routine,
and the associated amplitudes $g_W$ evaluated by using~(\ref{gfullres}).
The finite-size average amplitude $g_P^\avg$,
obtained as a flat average of the $2^P$ values of $g_W$ thus generated,
is shown in Figure~\ref{Iave} against period $P\le30$.
The last point involves $2^{30}=1\,073\,741\,824$ different games.
The plotted quantity oscillates as a function of the period.
These finite-size effects are however exponentially damped,
and so $g_P^\avg$ converges very fast to the asymptotic limit
$1/4$, consistently with (\ref{gavg}).

\begin{figure}[!ht]
\begin{center}
\includegraphics[angle=0,width=.8\linewidth]{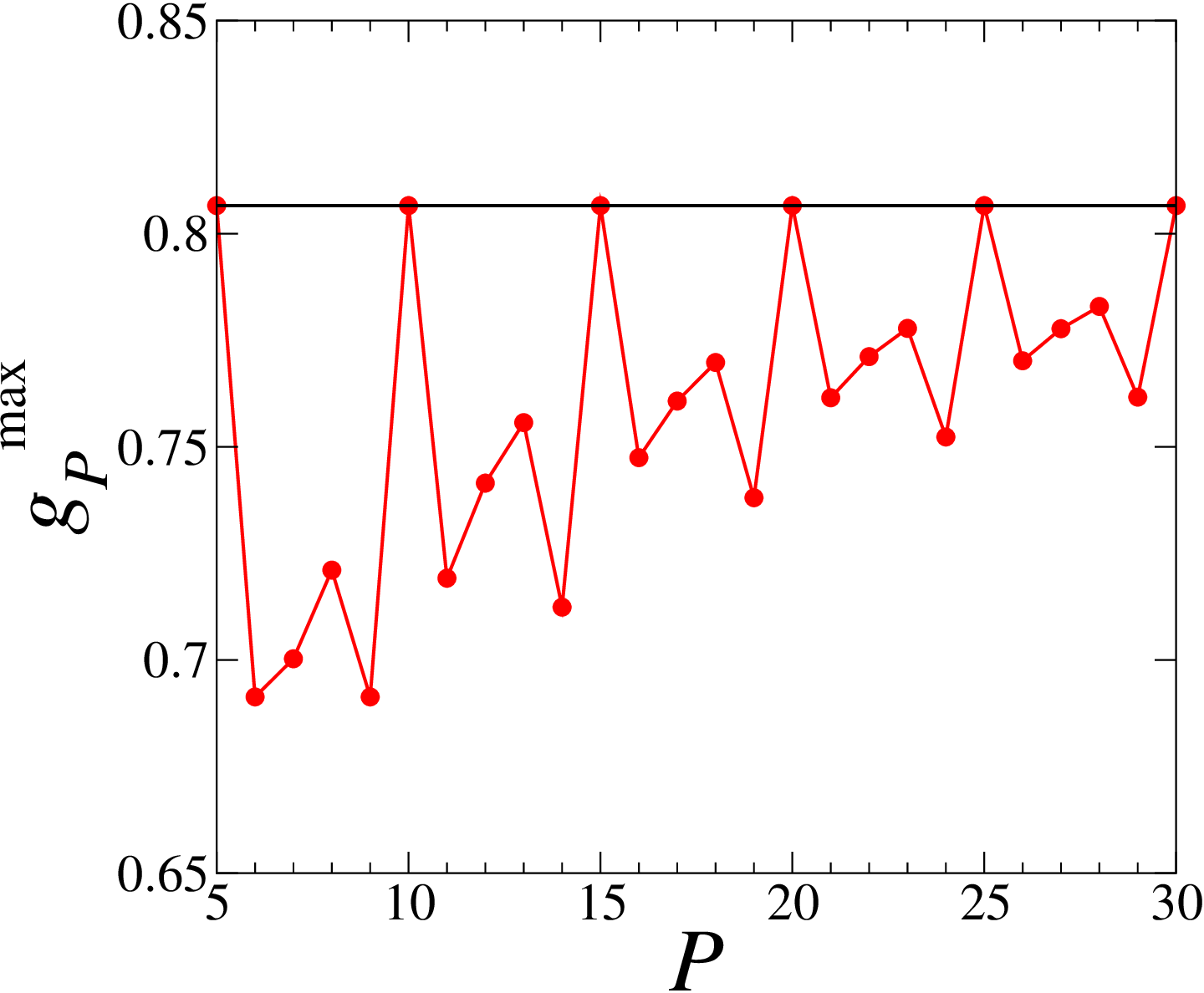}
\caption{
Maximal gain amplitude $g_P^\max$
of periodic capital-dependent Parrondo games with period $5\le P\le30$.
Horizontal black line: absolute maximum $g^\max$ (see~(\ref{gmaxI0})).}
\label{Imax}
\end{center}
\end{figure}

Let us now investigate which game yields the largest Parrondo effect,
i.e., the largest gain amplitude.
The maximal amplitude $g_P^\max$ among all $2^P$ periodic games with given period $P$
is shown in Figure~\ref{Imax} against $P$.
For the sake of clarity, the plotted range has been limited to $5\le P\le30$.
The amplitude of the periodic game with period~5 and unit cell $W=ABABB$,~i.e.,
\beq
g^\max=g_{ABABB}=\frac{488}{605}=0.806611
\label{gmaxI0}
\eeq
(see Table~\ref{pg}),
appears as the absolute maximum of the gain amplitudes of all games,
irrespective of their periods.
Whenever the period $P$ is a multiple of 5, the absolute maximum $g^\max$ is reached
for the game whose unit cell is a repetition of $p/5$ times $W$.
If $P$ is not a multiple of~5,
there are suboptimal periodic games
whose gains converge, albeit rather slowly, to (\ref{gmaxI0}).

We make a digression out of the weak-contrast regime
to mention that the periodic game $ABABB$
yields the highest gain for all values of the contrast parameter~$v$.
Its gain in the $v\to1$ limit, i.e.,
\beq
G^\max=G^\max_{ABABB}=\frac{9}{25}=0.36,
\label{gmaxI1}
\eeq
is the absolute maximal gain of the model in the neutral situation
where each rule, when played alone, is fair ($G_A=G_B=0$).
The $v\to1$ limit is however singular (see below~(\ref{rmaxI1})).
The universal optimality of the game $ABABB$
was already demonstrated by Dinis~\cite{LD}
by means of an algorithmic approach based upon backward induction.

It is interesting to notice that the values of $\r$
yielding the maximal gain of random games,
given by (\ref{rmaxI1}) for $v\to1$ and (\ref{rmaxI0}) for $v\to0$,
are very close to $3/5$, characteristic of the optimal periodic game $ABABB$.
The gains achieved by those optimal random games
are however far below the truly optimal values,
given by (\ref{gmaxI1}) for $v\to1$ and (\ref{gmaxI0}) for $v\to0$.

\subsection{Aperiodic games}
\label{Iwape}

The expression (\ref{gfullres}) for the gain amplitude extends to any aperiodic game,
either deterministic or random.
Taking formally the $P\to\infty$ limit,
forgetting about boundary conditions,
we obtain
\beq
g=\frac{2\rho}{3}-6\sum_{l,m=1}^\infty\q^{l+m}C_{l,m}.
\label{gaperes}
\eeq
In this expression,
\beq
\rho=\ave{\s_n}=\lim_{P\to\infty}\frac{1}{P}\sum_{n=1}^P\s_n
\eeq
is the density of letters $B$,
i.e., the fraction of steps where Rule~$B$ is chosen,
whereas
\beq
C_{l,m}=\ave{\s_n\s_{n+l}\s_{n-m}}
=\lim_{P\to\infty}\frac{1}{P}\sum_{n=1}^P\s_n\s_{n+l}\s_{n-m}
\eeq
are the three-point correlation functions of the distribution of letters $B$,
depending on two distances $l$ and $m$.
The damping factor $\q^{l+m}$ ensures an exponential convergence of (\ref{gaperes})
for all aperiodic games with well-defined translationally invariant correlations.

Hereafter we consider two examples of aperiodic games in more detail.
Games generated by chaotic dynamical systems
have already been considered in the past~\cite{TAA}.
The following examples are more directly inspired
by the physics of 1D systems.
The first example (Section~\ref{Irm})
consists of an enrichment of the random games considered in Section~\ref{Iwran}
by the introduction of a memory kernel.
The gain amplitude exhibits a smooth dependence on parameters (see Figure~\ref{Imark}).
The second example (Section~\ref{caps}) is based on quasiperiodic cut-and-project sequences.
The amplitude has an irregular dependence on parameters
(see Figure~\ref{Icut}).

\subsubsection{Random games with Markovian memory}
\label{Irm}

In Sections~\ref{Iran} and~\ref{Iwran} we have considered random games
where at each time step the rule is chosen at random,
irrespective of past and future.
In other words, the symbols~$\s_n$ introduced in (\ref{symbs})
are independent random variables.

The goal of this section is to consider a richer type of random games
based on random sequences with Markovian memory,
where at each step the rule is chosen with probabilities depending
on the rule at the previous step.
This setting allows two free parameters,
namely the probabilities $\a$ and $\b$, such that\footnote{Here and throughout the following,
w.~p.~is a shorthand for `with probability'.}
\beqa
&&\s_{n-1}=0\;\Rightarrow\;\s_n=\left\{
\matrix{0&\mbox{w.~p.~}\;1-\a,\cr 1&\mbox{w.~p.~}\;\a,\hfill}\right.
\nonumber\\
&&\s_{n-1}=1\;\Rightarrow\;\s_n=\left\{
\matrix{0&\mbox{w.~p.~}\;\b,\hfill\cr 1&\mbox{w.~p.~}\;1-\b.}\right.
\eeqa
In other words the game, i.e., the rule pattern, is generated by an auxiliary Markov chain,
whereas each rule, either $A$ or $B$, itself amounts to a Markov chain -- as before.
The above setting can be encoded into the Markov matrix
\beq
\m m=\pmatrix{1-\a & \b\cr \a & 1-\b}.
\eeq
The stationary state of the auxiliary Markov process
is described by the eigenvector $\m r$ such that
$\m r=\m m\m r$, i.e.,
\beq
\m r=\frac{1}{\a+\b}\pmatrix{\b\cr\a}.
\eeq
We have therefore
\beq
\rho=\frac{\a}{\a+\b}.
\eeq
The second eigenvalue of the Markov matrix $\m m$,
characterizing the range of the memory effect, reads
\beq
\l=1-\a-\b.
\eeq
In order to determine correlation functions,
an explicit representation of powers of $\m m$ is required.
We have
\beq
\m m^k=\pmatrix{1-\a_k & \b_k\cr \a_k & 1-\b_k},
\eeq
with $\a_{k+1}=\a+\l\a_k$ and $\b_{k+1}=\b+\l\b_k$, and so
\beq
\a_k=\rho(1-\l^k),\quad \b_k=(1-\rho)(1-\l^k).
\eeq

The Markovian property of the sequence defining the random game implies
\beq
C_{l,m}=\rho(1-\b_l)(1-\b_m).
\eeq
Inserting this expression into (\ref{gaperes}),
the double sum boils down to geometric series.
We are thus left with the explicit result
\beq
g=\frac{8\rho(1-\rho)(1-\l)((1-\l)\rho+2\l+1)}{3(2+\l)^2}.
\label{gmr}
\eeq

The amplitude vanishes as $\r\to0$ and $\r\to1$,
where random games respectively become Rule~$A$ and Rule~$B$.
The random games considered in Sections~\ref{Iran} and~\ref{Iwran}
correspond to an absence of memory, i.e., $\l=0$.
The result~(\ref{gres}) is thus recovered for the third time.

\begin{figure}[!ht]
\begin{center}
\includegraphics[angle=0,width=.75\linewidth]{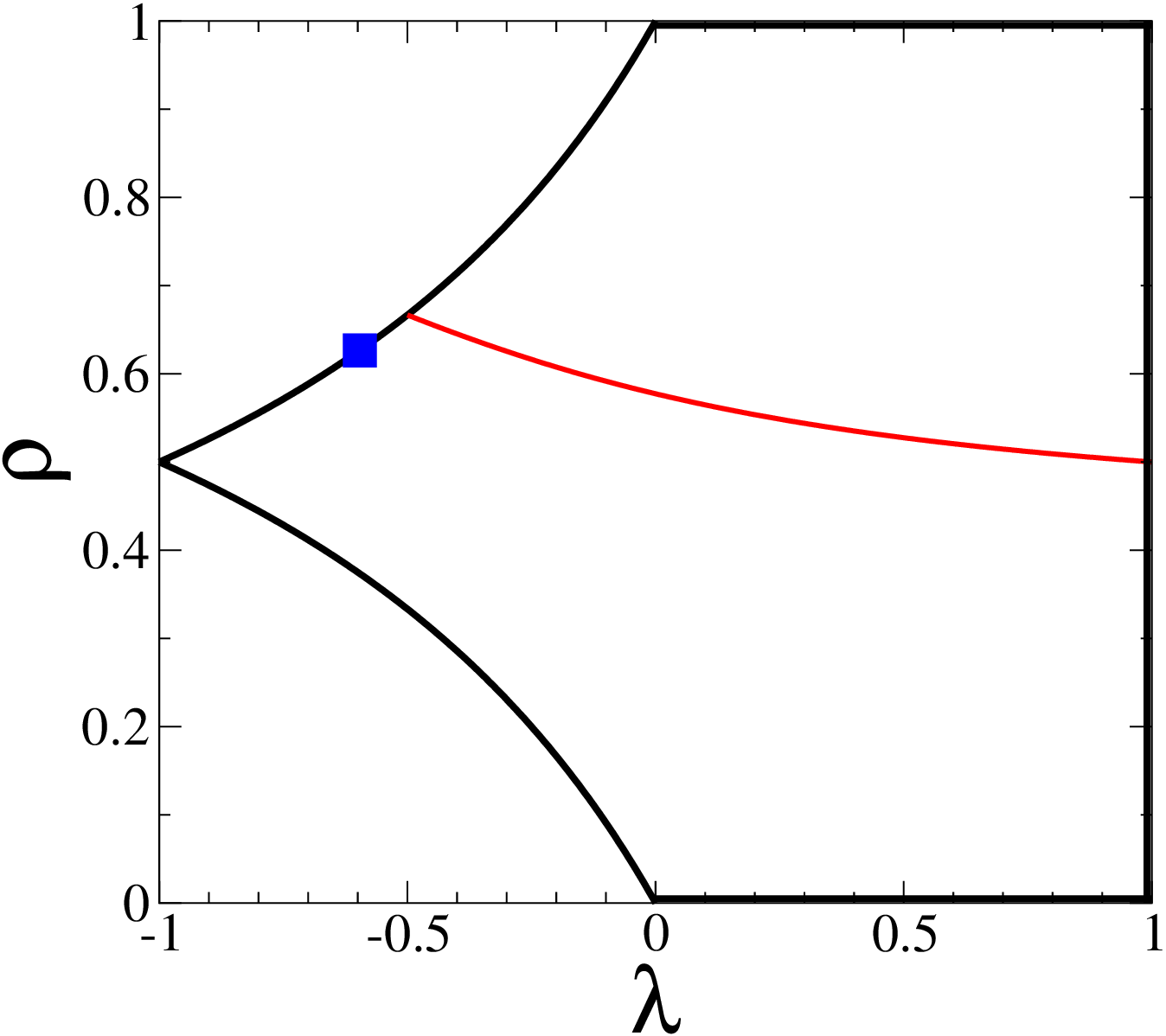}
\caption{
Parameter space of random sequences with Markovian memory
in the $(\l,\r)$ plane.
Allowed values lie inside the black curve.
Red curve: optimal density (see~(\ref{ropt})) where $g$ takes its maximum at fixed $\l$.
Blue square symbol: point where~$g$ takes its absolute maximum
(see (\ref{labs}), (\ref{rabs})).}
\label{Iphase}
\end{center}
\end{figure}

Figure~\ref{Iphase} shows the parameter space of random sequences with Markovian memory.
Allowed values of density $\r$ and memory rate $\l$ lie inside the black curve.
For $\l>0$, where successive symbols are positively correlated,
all values of the density $\r$ can be realized.
The gain vanishes linearly as $\l\to1$,
i.e., when the mean block length diverges.
For $\l<0$, where successive symbols are negatively correlated,
only a limited range of densities, i.e.,
\beq
-\frac{\l}{1-\l}\le\r\le\frac{1}{1-\l},
\label{lims}
\eeq
can be realized.
The upper (resp.~lower) bound corresponds to $\a=1$ (resp.~$\b=1$),
where letters $A$ (resp.~$B$) are isolated.
In the $\l\to-1$ limit, the range shrinks to the single point $\r=1/2$,
where the random game reduces to the periodic game $AB$.

\begin{figure}[!ht]
\begin{center}
\includegraphics[angle=0,width=.8\linewidth]{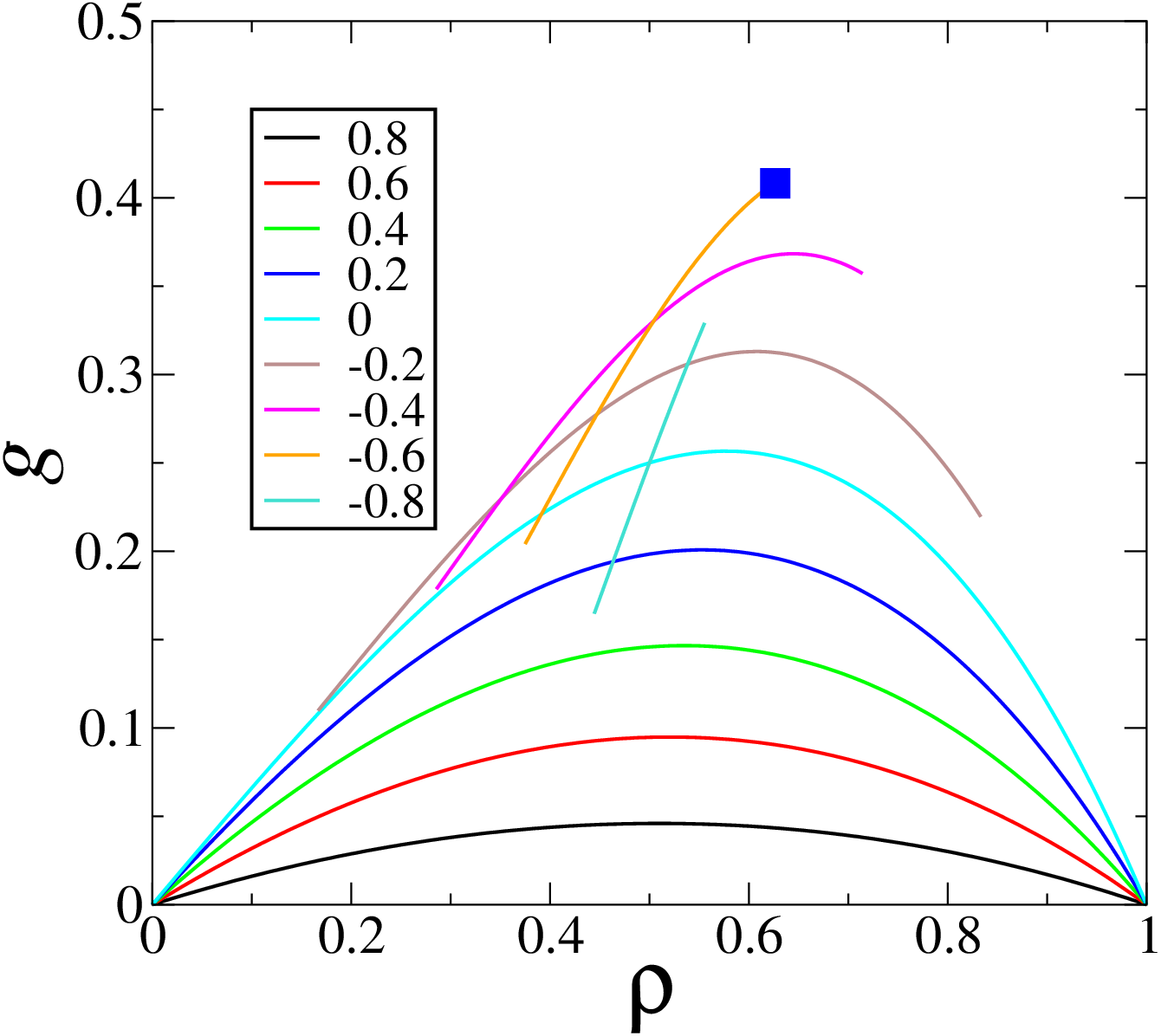}
\caption{
Dependence of the gain amplitude $g$ of random capital-dependent Parrondo games
with Markovian memory
on the density $\r$ of letters $B$, as given by (\ref{gmr}),
for several values of the memory rate $\l$ (see legend).
For negative $\l$, curves are limited to the range (\ref{lims}).
Blue square symbol: absolute maximum of~$g$ (see (\ref{gabs}), (\ref{rabs})).}
\label{Imark}
\end{center}
\end{figure}

Figure~\ref{Imark} shows the dependence of the amplitude~$g$ on the density $\r$
of letters $B$, as given by (\ref{gmr}), for several values of the memory rate $\l$.

For fixed $\l$, $g$ reaches its maximum
\beq
g_\max=\frac{8\left(2\sqrt{3}(1+\l+\l^2)^{3/2}-9\l(1+\l)\right)}{27(1-\l)(2-\l)^2}
\eeq
for
\beq
\r=\frac{\sqrt{3}(1+\l+\l^2)^{1/2}-3\l}{3(1-\l)}.
\label{ropt}
\eeq
The dependence of this optimal density on $\l$
is shown in Figure~\ref{Iphase} as a red curve.
The latter leaves the range of allowed densities
as it hits the $\a=1$ boundary for $\b=1/2$, i.e., $\l=-1/2$ and $\r=2/3$,
where $g=32/81=0.395061$.

The gain amplitude however reaches a slightly higher absolute maximum,
\beqa
g_\max&=&\frac{2}{9}\left((16\sqrt{2}-13)^{1/3}-7(16\sqrt{2}-13)^{-1/3}+3\right)
\nonumber\\
&=&0.408187,
\label{gabs}
\eeqa
somewhere further along the $\a=1$ boundary, i.e., for
\beqa
\l&=&(1+\sqrt{2})^{-1/3}-(1+\sqrt{2})^{1/3}
\nonumber\\
&=&-0.596071,
\label{labs}
\\
\r&=&\frac{1}{6}\left((8+6\sqrt{2})^{1/3}-2(8+6\sqrt{2})^{-1/3}+2\right)
\nonumber\\
&=&0.626538.
\label{rabs}
\eeqa
This optimal point is shown as blue square symbols in Figures~\ref{Iphase} and~\ref{Imark}.

\subsubsection{Cut-and-project quasiperiodic games}
\label{caps}

Our second example of aperiodic games is very different in spirit.
It is generated by the deterministic quasiperiodic cut-and-project sequences.
These sequences, investigated first by de Bruijn~\cite{deb},
are in correspondence with irrational numbers~$\o$.
They have been extensively used to build model quasiperiodic structures
that are 1D analogues of quasicrystals.
In particular,
for $\o=1/\tau$ and $\o=1/\tau^2$,
where $\tau=(1+\sqrt{5})/2=1.618033$ is the golden mean,
Fibonacci sequences are obtained,
which are germane to the first icosahedral quasicrystals,
discovered in~1984~\cite{she}
(see~\cite{janot,sene} for overviews).
Since then, much attention has been paid
to cut-and-project and other deterministic aperiodic sequences
and to various physical models based upon these structures
(see~\cite{albu,macia} for reviews).

The cut-and-project sequence
is based on an irrational rotation number in the range $0<\o<1$.
Consider the points obtained by rotating around the unit circle
in discrete steps by the angle $\o$, measured in revolutions, i.e., in units of $2\pi$.
The angle reached after $n$ steps reads
\beq
x_n=\Frac(n\o),
\eeq
where $\Frac(x)=x-\Int(x)$ is the fractional part of a real number $x$,
with $\Int(x)$ being its integer part.
The binary cut-and-project sequence of symbols $\s_n$ is defined by setting
\beq
\s_n=\chi(x_n),
\eeq
where
\beq
\chi(x)=\left\{
\matrix{1\quad&(0\le x<\o),\hfill\cr 0\hfill&(\o\le x<1).}\right.
\eeq
In other words, we have $\s_n=1$ if the angle $x_n$ is in the interval $[0,\,\o[$,
and $\s_n=0$ otherwise.

We consider the infinitely long
Parrondo game defined by choosing Rule~$A$ (resp.~Rule~$B$)
at step $n$ if $\s_n=0$ (resp.~$\s_n=1$),
consistently with (\ref{symbs}).
For all irrational rotation numbers $\o$,
the sequence $x_n$ is uniformly distributed over $[0,\;1]$,
so that the density of letters $B$,
i.e., the fraction of steps where Rule~$B$ is chosen,
reads
\beq
\r=\o.
\label{cutr}
\eeq
The fluctuations in the letter numbers,
measured by the differences
\beq
\delta_n=\sum_{m=1}^n\s_m-n\o,
\label{flucs}
\eeq
belong to the interval $-1\le\delta_n\le0$.
They are therefore bounded,
whereas they would typically grow as $\sqrt{n}$ for a random sequence.

\begin{figure}[!ht]
\begin{center}
\vskip -10pt
\includegraphics[angle=0,width=.75\linewidth]{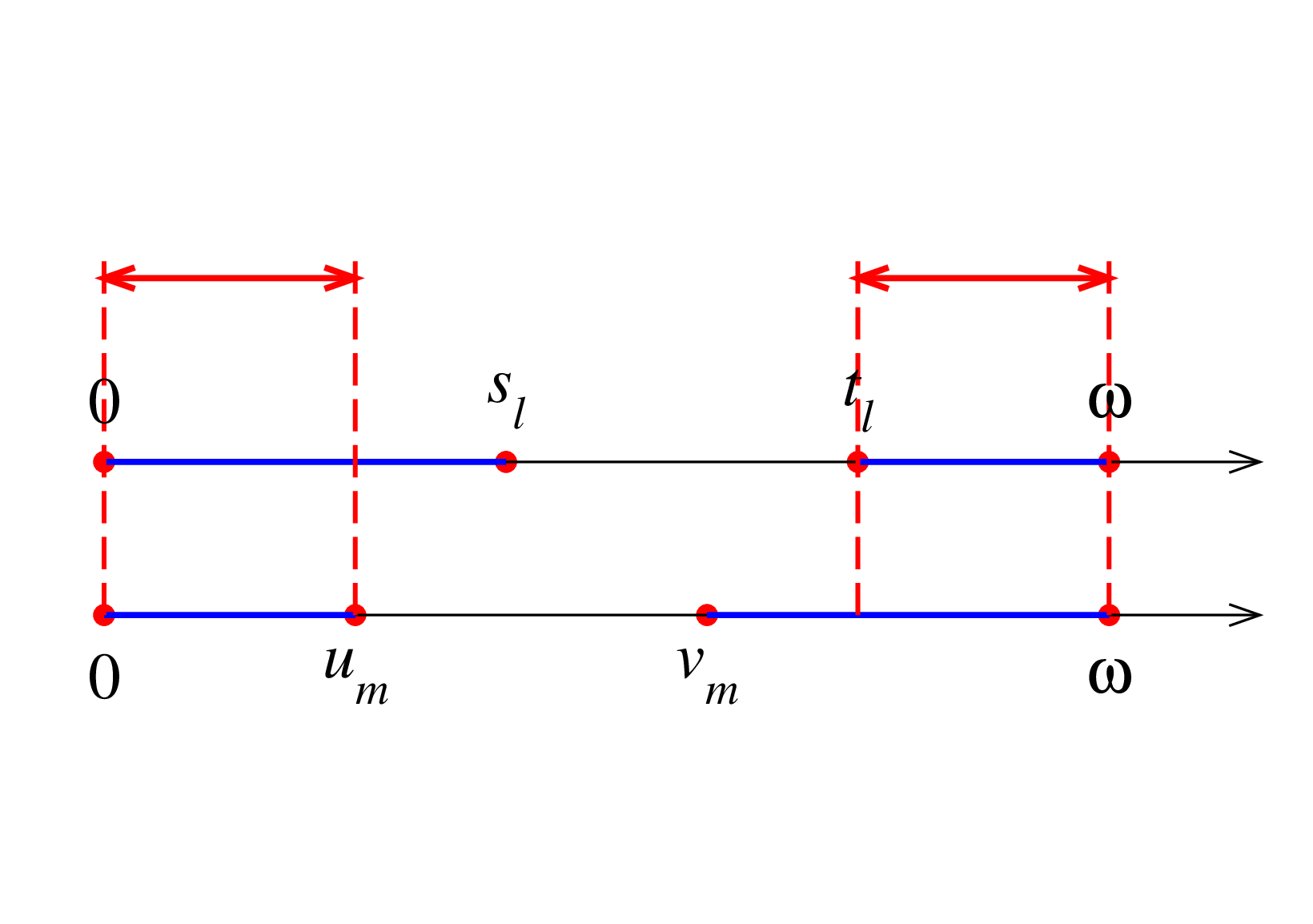}
\caption{
Construction of the set involved in the determination
of the three-point correlation function $C_{l,m}$
of the cut-and-project sequence.
The latter quantity is the length of the intervals marked by red arrows,
defined as the intersection of the blue sets drawn on each axis.
Notations are given in (\ref{nots}).}
\label{Iinter}
\end{center}
\end{figure}

The correlation function $C_{l,m}$
is given by the length of the set of values of $x$
such that the three numbers $x$, $\Frac(x+l\o)$ and $\Frac(x-m\o)$
all belong to $[0,\;\o]$.
The construction of this set is sketched in Figure~\ref{Iinter},
with the notations
\beqa
&&s_l=\max(\o-x_l,0),\quad t_l=\min(1-x_l,\o),
\nonumber
\\
&&u_m=\max(\o-1+x_m,0),\quad v_m=\min(x_m,\o).
\label{nots}
\eeqa

The expression
\beqa
C_{l,m}&=&\min(u_m,s_l)+\o-\max(t_l,v_m)
\nonumber\\
&+&\max(s_l-v_m,0)+\max(u_m-t_l,0)
\label{cutclm}
\eeqa
synthesizes the six different possible orders
between the four points $s_l$, $t_l$, $u_m$ and $v_m$
(we have always $s_l<t_l$ and $u_m<v_m$).

\begin{figure}[!ht]
\begin{center}
\includegraphics[angle=0,width=.8\linewidth]{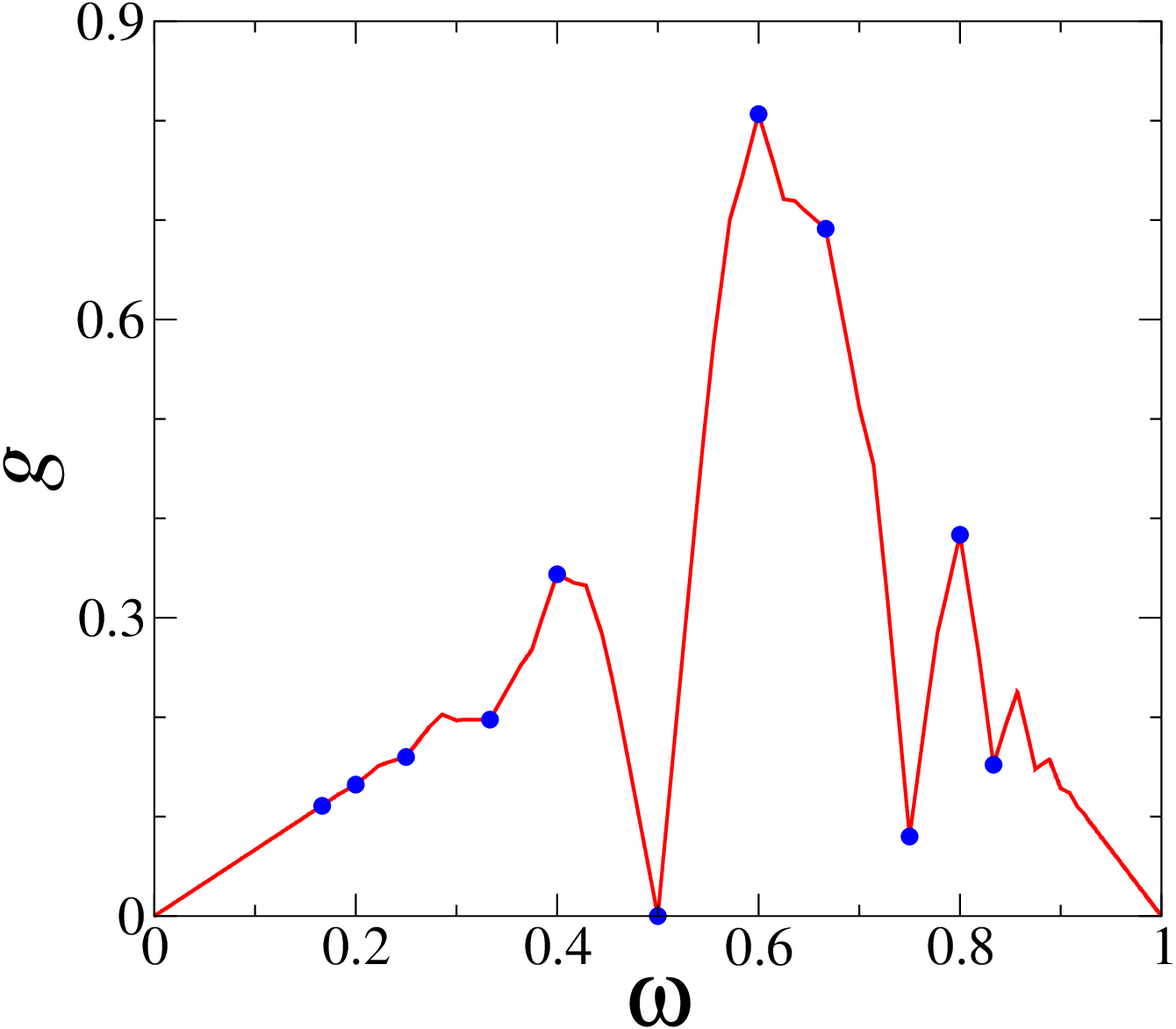}
\vskip 6pt
\caption{
Dependence of the gain amplitude $g$
for the capital-depen\-dent Parrondo game
against the rotation number $\o$ defining the cut-and-project sequence.
Blue symbols: rational rotation numbers with denominator $P\le6$
(see Table~\ref{pg}).}
\label{Icut}
\end{center}
\end{figure}

Figure~\ref{Icut} shows the gain amplitude~$g$
against the rotation number $\o$ of the cut-and-project game,
as obtained by inserting the expressions (\ref{cutr}) and (\ref{cutclm}) into (\ref{gaperes}),
evaluating individual terms and performing the sum numerically.

The amplitude $g$ appears to be a continuous function of~$\o$,
exhibiting cusps at rational values of $\o$,
around which it varies linearly,
albeit with two different slopes to the left and to the right.
If $\o$ goes to a rational $Q/P$, assumed irreducible,
the corresponding sequence becomes periodic, with period~$P$.
Only a very specific subset of periodic sequences is attained in this way.
The last column of Table~\ref{pg} gives the values of~$\o$ corresponding
to all periodic games thus obtained with primitive periods $P\le6$.
The corresponding data points are shown as blue symbols in Figure~\ref{Icut}.
The amplitude vanishes only for $\o=0$ (Rule~$A$), $\o=1$ (Rule~$B$)
and $\o=1/2$ (periodic game $AB$).
It reaches its maximum (see~(\ref{gmaxI0})) for $\o=3/5$.

The amplitude vanishes linearly in the vicinity of both endpoints
($\o\to0$ and $\o\to1$),
up to exponentially small deviations.
For $\o\to0$,
the smallest distances yielding a non-zero
three-point correlation function $C_{l,m}$
are $m=l-1=\Int(1/\o)$.
A similar line of reasoning applies to $\o\to1$ as well.
We thus obtain the estimates
\beqa
g&=&\frac{2\o}{3}+O(2^{-2/\o})\quad(\o\to0),
\\
g&=&\frac{4(1-\o)}{3}+O(2^{-1/(1-\o)})\quad(\o\to1).
\eeqa

\section{History-dependent games}
\label{II}

\subsection{Generalities}
\label{IIgen}

We now turn to history-dependent Parrondo games~\cite{PHA,KJ,EL}.
In this second class of games,
the walker moves either right or left at step $t$,
i.e., its $t$th step
\beq
\eps_t=n_t-n_{t-1}
\eeq
is chosen to be either $\eps_t=+1$ or $\eps_t=-1$,
with probabilities which are independent of its position $n_t$,
but depend on the~$Q$ previous steps,
in a way that is different for Rules~$A$ and $B$.

Hereafter we restrict the analysis to the smallest relevant memory range, i.e., $Q=2$.
It is sufficient to characterize the system by the four-dimensional
time-dependent state vector
\beq
\m\phi_t=\pmatrix{X_t\cr Y_t\cr Z_t\cr T_t},
\eeq
with
\beqa
X_t&=&\prob\,(\eps_{t-1}=+1\mbox{ and }\eps_t=+1),
\nonumber\\
Y_t&=&\prob\,(\eps_{t-1}=+1\mbox{ and }\eps_t=-1),
\nonumber\\
Z_t&=&\prob\,(\eps_{t-1}=-1\mbox{ and }\eps_t=+1),
\nonumber\\
T_t&=&\prob\,(\eps_{t-1}=-1\mbox{ and }\eps_t=-1).
\eeqa
The mean displacement during the $t$th step reads
\beq
\mean{\eps_t}=X_t-Y_t+Z_t-T_t=\m J\cdot\m\phi_t,
\eeq
where the displacement vector reads
\beq
\m J=\pmatrix{1&-1&1&-1}.
\eeq
The expression~(\ref{Gdef}) of the gain therefore translates to
\beq
G=\lim_{t\to\infty}\frac{1}{t}\sum_{s=1}^t\m J\cdot\m\phi_s.
\label{IIG}
\eeq

The usual class of history-dependent Parrondo games
consists of a combination of the following rules~\cite{PHA,KJ,EL}.

\begin{itemize}

\item[$\bullet$] Rule~$A$.
This rule coincides with Rule~$A$ in capital-dependent games.
In the present setting,
each step is chosen according to
\beq
\eps_t=\left\{
\matrix{+1&\mbox{w.~p.~}\;p,\hfill\cr -1&\mbox{w.~p.~}\;q=1-p,}\right.
\eeq
irrespective of the past,
where the notation $p$ is consistent with Sections~\ref{I} and~\ref{Iweak}.
Therefore, if Rule~$A$ is played at time $t$, we have
\beq
\m\phi_t=\m M_A\m\phi_{t-1},
\eeq
with
\beq
\m M_A=\pmatrix{p&0&p&0\cr q&0&q&0\cr 0&p&0&p\cr 0&q&0&q}.
\eeq
If Rule~$A$ is played alone,
the walker executes a uniformly biased random walk.
Its stationary state reads
\beq
\m\phi_A=\pmatrix{p^2\cr pq\cr pq\cr q^2}.
\eeq
We have (see~(\ref{IIG}))
\beq
G_A=\m J\cdot\m\phi_A,
\eeq
i.e.,
\beq
G_A=2p-1,
\eeq
consistently with~(\ref{IGA}).

\item[$\bullet$] Rule~$B$.
This is the most general rule with memory range $Q=2$,
If Rule~$B$ is played at time~$t$,
the displacement $\eps_t=\pm1$ is chosen according to the following stochastic rules,
depending on the two previous steps $(\eps_{t-2},\eps_{t-1})$:
\beqa
(+1,+1)&\Rightarrow&
\eps_t=\left\{\matrix{+1&\mbox{w.~p.~}\;p_1,\cr -1&\mbox{w.~p.~}\;q_1,\hfill}\right.
\nonumber\\
(+1,-1)&\Rightarrow&
\eps_t=\left\{\matrix{+1&\mbox{w.~p.~}\;p_2,\cr -1&\mbox{w.~p.~}\;q_2,\hfill}\right.
\nonumber\\
(-1,+1)&\Rightarrow&
\eps_t=\left\{\matrix{+1&\mbox{w.~p.~}\;p_3,\cr -1&\mbox{w.~p.~}\;q_3,\hfill}\right.
\nonumber\\
(-1,-1)&\Rightarrow&
\eps_t=\left\{\matrix{+1&\mbox{w.~p.~}\;p_4,\cr -1&\mbox{w.~p.~}\;q_4,\hfill}\right.
\eeqa
with the notation $q_i=1-p_i$.
The $p_i$ are considered as four free parameters.

We have therefore
\beq
\m\phi_t=\m M_B\m\phi_{t-1},
\eeq
with
\beq
\m M_B=\pmatrix{p_1&0&p_3&0\cr q_1&0&q_3&0\cr 0&p_2&0&p_4\cr 0&q_2&0&q_4}.
\eeq

If Rule~$B$ is played alone,
the stationary state of the system reads
\beq
\m\phi_B=\pmatrix{X_B\cr Y_B\cr Z_B\cr T_B},
\eeq
with
\beq
X_B=\frac{p_3p_4}{D},\quad Y_B=Z_B=\frac{q_1p_4}{D},\quad T_B=\frac{q_1q_2}{D}
\eeq
and
\beq
D=q_1q_2+2q_1p_4+p_3p_4.
\eeq
We have (see~(\ref{IIG}))
\beq
G_B=\m J\cdot\m\phi_B,
\eeq
i.e.,
\beq
G_B=\frac{p_3p_4-q_1q_2}{D}.
\label{IIGB}
\eeq

\end{itemize}

Hereafter the main focus will again be on the neutral situation
where each rule, when played alone, is fair ($G_A=G_B=0$).
The condition for Rule~$A$ to be fair is again~(\ref{afair}),
expressing that the corresponding random walk is symmetric.
The condition that Rule~$B$ is fair reads
\beq
p_3p_4=q_1q_2.
\eeq
This non-linear relation leaves three free parameters.
We choose the parame\-trization
\beq
q_1=\frac{ab}{c},\quad
q_2=\frac{ac}{b},\quad
p_3=\frac{a}{bc},\quad
p_4=abc,
\label{pqabc}
\eeq
and introduce for further convenience the logarithmic co-ordinates
\beq
a=\e^{-\lambda},\quad b=\e^u,\quad c=\e^v.
\label{abcluv}
\eeq

\begin{figure}[!ht]
\begin{center}
\includegraphics[angle=0,width=.8\linewidth]{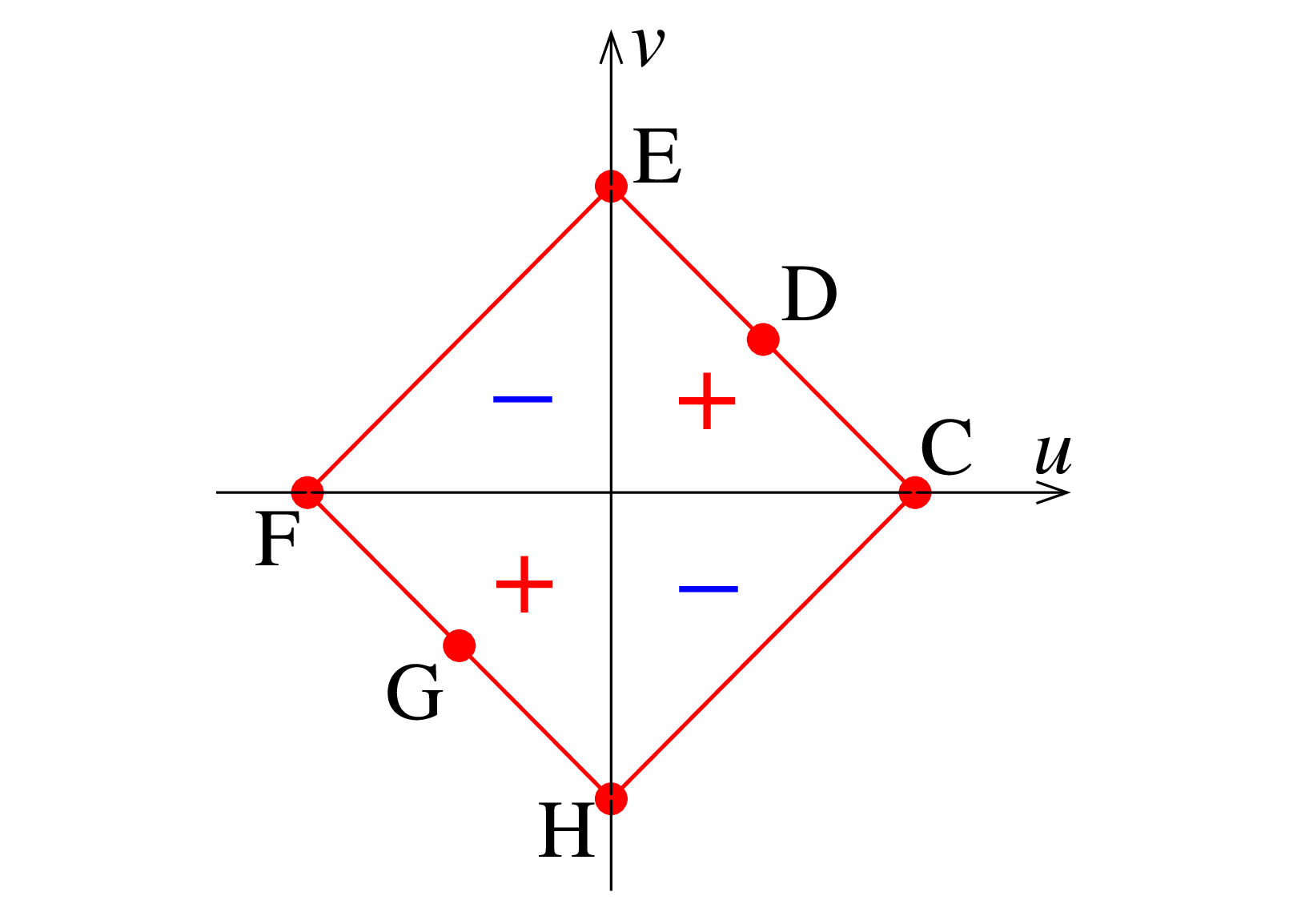}
\caption{
Parameter space of Rule~$B$
of the history-dependent Parrondo game
in the ($u,v$) plane
for a fixed value of $a$ in the range $0<a<1$.
Allowed parameters lie inside the square (see text).}
\label{IIefgh}
\end{center}
\end{figure}

Figure~\ref{IIefgh} shows the parameter space of the neutral situation
in the ($u,v$) plane,
for a fixed value of $a$ in the range $0<a<1$.
Allowed parameter values lie inside a square
with vertices C$(\lambda,0)$, E$(0,\lambda)$, F$(-\lambda,0)$ and H$(0,-\lambda)$.
The edges of the square correspond to limiting cases:
we have $p_4=1$ along CE, $q_2=1$ along EF, $p_3=1$ along FH and $q_1=1$ along HC.
Symbols $+$ and $-$ refer to the sign of the gain (see below~(\ref{delave})).
The midpoints D ($q_1=q_2=a$, $p_3=a^2$, $p_4=1$)
and G ($q_1=q_2=a$, $p_3=1$, $p_4=a^2$)
of the edges CE and FH
play a part in the subsequent discussion.

Parity, i.e., the change of sign of the walker's position ($n\lra -n$),
amounts to exchanging parameters according to $p\lra q$ for Rule~$A$,
and for Rule~$B$ $p_4\lra q_1$, $p_3\lra q_2$, i.e., $c\lra1/c$ or $v\lra-v$.
Parity therefore amounts to a reflection of Figure~\ref{IIefgh}
with respect to its horizontal $u$-axis.
No symmetry is associated with the reflection of Figure~\ref{IIefgh}
with respect to its vertical $v$-axis.
Moreover, at variance with the capital-dependent games considered in Sections~\ref{I} and~\ref{Iweak},
the history-dependent Parrondo games considered here
do not exhibit any simple transformation under time reversal.

\subsection{Random games}
\label{IIran}

The first situation of interest demonstrating Parrondo's paradox is again that of random games,
where at each time step Rule~$B$ is chosen with probability $\r$
and Rule~$A$ with the complementary probability $1-\r$.
In order to determine the average gain $\ave{G}$ of random games,
it is sufficient to know the stationary average state vector $\ave{\m\phi}$.
The average Markov matrix $\ave{\m M}$ (see~(\ref{avem}))
again has the same functional form as $\m M_B$,
with effective parameters
\beqa
\ave q_1&=&(1-\r)q+\r q_1,\quad \ave q_2=(1-\r)q+\r q_2,
\nonumber\\
\ave p_3&=&(1-\r)p+\r p_3,\quad \ave p_4=(1-\r)p+\r p_4.
\eeqa
The average gain $\ave G$ is obtained by
replacing all parameters entering (\ref{IIGB}) by the above effective values.

For the uniformly random game ($\r=1/2$),
where at each time step Rules~$A$ and $B$ are chosen with equal probabilities,
we thus obtain
\beq
\ave G=\frac{p_3p_4-q_1q_2+p(p+p_3+p_4)-q(q+q_1+q_2)}{D},
\label{IIgur}
\eeq
with
\beqa
D&=&q_1q_2+2q_1p_4+p_3p_4
\nonumber\\
&+&p(p+p_3+p_4+2q_1)
\nonumber\\
&+&q(q+q_1+q_2+2p_4).
\eeqa
The expression (\ref{IIgur}) again allows one to measure the rarity of Parrondo's paradox.
We define the probability of observing Parrondo's paradox
as the volume of the five-dimensional domain in $(p,p_1,p_2,p_3,p_4)$ space such that
the inequalities (\ref{pdef}) hold, with $G$ given by~(\ref{IIgur}).
A numerical integration again yields a very small number (see~(\ref{pc}))
\beq
\prob(\hbox{Parrondo's paradox})\approx0.000505.
\label{ph}
\eeq

From now on,
we restrict the analysis to history-depen\-dent Parrondo games in the neutral situation
where each rule, when played alone, is fair ($G_A=G_B=0$).
Using the parametri\-zation (\ref{afair}), (\ref{pqabc}),
we obtain the following expression for the average gain:
\beq
\ave{G}=\frac{a\r(1-\r)(b^2-1)(c^2-1)}{D},
\label{IIaveg}
\eeq
with
\beqa
D&=&2(1-\r)^2bc+\r(1-\r)a(3b^2+1)(c^2+1)
\nonumber\\
&+&4\r^2a^2bc(b^2+1).
\label{delave}
\eeqa

The expression (\ref{IIaveg}) shows that the gain has the sign of the product $(b^2-1)(c^2-1)$,
i.e., equivalently, of the product $uv$,
irrespective of $a$ and of the probability $\r$.
Therefore, in the neutral situation under consideration,
Parrondo's paradox holds in one half of parameter space,
i.e., in the two regions marked by $+$ signs in Figure~\ref{IIefgh}.
The average gain vanishes as $\r\to0$ and $\r\to1$,
where random games respectively degenerate to Rule~$A$ and Rule~$B$.
It reaches its absolute maximum,
\beq
\ave{G}^\max\to1,
\label{avetriv}
\eeq
in the limit where $a\to0$ and $\r\to1$ simultaneously.
More precisely, for a fixed small value of~$a$,
the average gain $\ave{G}$ reaches its maximum with respect to $\r$, $b$ and $c$ for
\beq
\r\approx1-\sqrt{2}\,a,\quad
b\approx2^{-1/4}\sqrt{a},\quad c\approx2^{1/4}\sqrt{a}.
\eeq
The corresponding point in Figure~\ref{IIefgh} is along the edge FH
and close to its midpoint~G.
This maximum reads
\beq
\ave{G}^\max(a)\approx1-8\sqrt{2}\,a,
\eeq
so that (\ref{avetriv}) is attained in the $a\to0$ limit.
This limit is however singular
-- irrespective of the parameters $b$ and~$c$,
provided they remain in the allowed range --
as another eigenvalue of the Markov matrix $\m M_B$ goes to unity,
so that the latter matrix loses its property of unique ergodicity.

The weak-contrast scaling regime is
defined by the conditions that both parameters~$b$ and $c$ are close to unity,
i.e., that $u$ and $v$ are simultaneously small.
This scaling regime therefore corresponds to zooming on the center of Figure~\ref{IIefgh}.
At variance with the situation of capital-dependent games,
in the present case the weak-contrast regime keeps one free parameter, $a$.
For random games,
the expression~(\ref{IIaveg}) for the average gain vanishes proportionally to $uv$.
We shall see in Section~\ref{IIweak} that a similar scaling holds for arbitrary games.
We are thus led to introduce the gain amplitude
\beq
g=\lim_{u,v\to0}\frac{G}{uv}.
\label{IIgdef}
\eeq
For random games, (\ref{IIaveg}) yields
\beq
\ave{g}=\frac{2a\r(1-\r)}{(1+(2a-1)\r)^2}.
\label{IIgres}
\eeq
For the uniformly random game ($\r=1/2$),
this reads
\beq
\ave{g}=\frac{2a}{(1+2a)^2}.
\label{IIgavg}
\eeq
When the probability $\r$ of choosing Rule~$B$ varies between~0 and 1,
the amplitude (\ref{IIgres}) reaches its maximum
\beq
\ave{g}=\frac{1}{4}
\eeq
for
\beq
\r=\frac{1}{1+2a}.
\eeq

\subsection{Periodic games}
\label{IIper}

We now turn to periodic games,
defined by the periodic repetition of a unit cell $W$ of length $P$.
Here, too, the stationary state of the game has the same period~$P$ as the game itself.
It is encoded in $P$ state vectors $\m\phi_n$ obeying
\beq
\m\phi_n=\m M_{\tau_n} \m\phi_{n-1}\quad(n=1,\dots,P),
\label{IIeqper}
\eeq
with the notation (\ref{symbs}),
and with periodic boundary conditions ($\m\phi_P=\m\phi_0$).
The associated gain reads
\beq
G_W=\frac{1}{P}\sum_{n=1}^P\m J\cdot\m\phi_n
\eeq
(see (\ref{IIG})).
The recursion (\ref{IIeqper}) amounts to a system of~$4P$ linear equations.
The complexity of the expressions of the gain $G$ again grows very rapidly with the period $P$.
The gain is invariant under cyclic permutations,
but not under reversal of the unit cell.
Its expressions for periods 2 and~3 are as follows.

\begin{itemize}

\item[$\bullet$] $P=2$.
There is only one unit cell with period 2.
The corresponding gain reads
\beq
G_{AB}=\frac{(b^2-1)(c^2-1)}{2(b^2+1)(c^2+1)}.
\eeq

\item[$\bullet$] $P=3$.
There are two unit cells with period 3.
The corresponding gains read
\beqa
G_{AAB}&=&\frac{a(b^2-1)(c^2-1)}{6bc},
\\
G_{ABB}&=&\frac{a(b^2-1)(c^2-1)}{3b}\,\frac{N}{D},
\eeqa
with
\beqa
N&=&3b^2c-ab(2b^2+1)(c^2+1)+a^2c(b^2+1)^2,
\nonumber\\
D&=&2b^2c^2+abc(b^2-1)(c^2+1)
\nonumber\\
&-&a^2(b^2+1)(b^2-2c^2+b^2c^4).
\eeqa

\end{itemize}

The above expressions demonstrate that the gain vanishes
proportionally to $(b^2-1)(c^2-1)$, i.e., to $uv$
in the weak-contrast regime.
The corresponding gain amplitudes (see~(\ref{IIgdef}))
are listed in the first three lines of Table~\ref{IIpg}.

\section{Weak-contrast scaling regime of history-dependent games}
\label{IIweak}

\subsection{Generalities}
\label{IIwgen}

The problem again simplifies in the weak-contrast regime ($u,v\to0$).
Hereafter we use the shorthand notation
\beq
\mu=1-2a,
\eeq
so that $0<a<1$ translates to $\abs{\mu}<1$.

For the periodic games considered in Section~\ref{IIper},
the matrix recursion (\ref{IIeqper}) boils down
to two coupled linear recursion relations for the rescaled co-ordinates
\beqa
x_n&=&1+\lim_{u,v\to0}\frac{X_n-Y_n-Z_n+T_n}{u},\quad
\\
y_n&=&\lim_{u,v\to0}\frac{X_n-Y_n+Z_n-T_n}{uv},
\eeqa
namely, with the notation (\ref{symbs}):
\beqa
\s_n&=&0\;\Rightarrow\;\left\{\matrix{x_n=1,\hfill\cr y_n=0,\hfill}\right.
\label{II0r}
\\
\s_n&=&1\;\Rightarrow\;\left\{\matrix{x_n=\mu x_{n-1},\hfill\cr y_n=\mu y_{n-2}+(1-\mu)x_{n-1},\hfill}\right.
\label{II1r}
\eeqa
with periodic boundary conditions ($y_P=y_0$, $x_P=x_0$).
The gain amplitude (see~(\ref{IIgdef})) reads
\beq
g_W
=\lim_{u,v\to0}\frac{G_W}{uv}
=\frac{1}{P}\sum_{n=1}^Py_n.
\label{IIgsum}
\eeq
Here, too, the above formalism extends to aperiodic games (see Section~\ref{IIwape}).

There are analogies and differences between the studies of the weak-contrast regimes
exposed in Sections~\ref{Iwgen} and~\ref{IIwgen}.
The main difference is that in (\ref{Iyr}), (\ref{Ixr})
the damping factor $\q$ is uniform
and the variable $\s_n$ encoding the rule applied at step $n$ enters linearly,
whereas the full structure of the recursions (\ref{II0r}), (\ref{II1r}) depends on~$\s_n$.

\subsection{Random games}
\label{IIwran}

As a first application of the above formalism,
let us revisit random games, considered in Section~\ref{IIran}.
As a consequence of (\ref{II0r}), (\ref{II1r}),
the stationary averages~$\ave{x}$ and $\ave{y}$ obey
\beqa
\ave{x}&=&1-\r+\mu\r\ave{x},
\\
\ave{y}&=&\r(\mu\ave{y}+(1-\mu)\ave{x}),
\eeqa
hence
\beqa
\ave{x}&=&\frac{1-\r}{1-\mu\r},
\\
\ave{g}&=&\ave{y}=\frac{(1-\mu)\r(1-\r)}{(1-\mu\r)^2}.
\eeqa
The result (\ref{IIgres}) is thus recovered.

\subsection{Periodic games}
\label{IIwp}

We now revisit the situation of periodic games,
considered in Section~\ref{IIper}.
At variance with (\ref{Iyr}), (\ref{Ixr}),
where the variable $\s_n$ enters linearly,
allowing for the explicit solution (\ref{gfullres}),
in the present situation (\ref{II0r}), (\ref{II1r})
cannot be solved in closed form for periodic games with arbitrary unit cell~$W$.

An explicit formula for the gain amplitude can however be obtained
in the case where the unit cell consists of only two blocks (see~(\ref{wab})), i.e.,
\beq
W=A^MB^N,
\eeq
with $M$, $N\ge1$ and $P=M+N$.
The form of the result depends on whether~$M$ is one or larger,
and on the parity of $N$.
Omitting details, we obtain
\beqa
&&{\hskip -50pt}\bullet M=1,\ N=2k:
\nonumber\\
&&g_{A^MB^N}=\frac{(1-\mu^k)(1-\mu^{2k+1})}{P(1-\mu)},
\label{IIab1}
\\
&&{\hskip -50pt}\bullet M=1,\ N=2k+1:
\nonumber\\
&&g_{A^MB^N}=\frac{1-\mu^{k+1}}{P(1-\mu)},
\\
&&{\hskip -50pt}\bullet M\ge2,\ N=2k:
\nonumber\\
&&g_{A^MB^N}=\frac{(1-\mu^k)(1-\mu^{k+1})}{P(1-\mu)},
\\
&&{\hskip -50pt}\bullet M\ge2,\ N=2k+1:
\nonumber\\
&&g_{A^MB^N}=\frac{(1-\mu^{k+1})^2}{P(1-\mu)}.
\label{IIab4}
\eeqa
When both blocks lengths $M$ and $N$ become simultaneously large,
the amplitude falls off as
\beq
g_{A^MB^N}\approx\frac{1}{P(1-\mu)}=\frac{1}{2aP},
\eeq
up to exponentially small corrections.
This $1/P$ fall-off can again be interpreted
by stating that only the interfaces between blocks yield some gain.

We now turn to general features of interest
exhibited by the amplitudes of periodic games.
The dependence of the amplitude $g_W$ on the unit cell $W$
again appears to be very intricate in general.
Table~\ref{IIpg} gives the product $Pg_W$ for all periodic games with
primitive period $P\le6$.
The explicit results (\ref{IIab1})--(\ref{IIab4}) yield 15 of the 21 expressions given there,
whereas the remaining six cases require a specific solution
of the recursion (\ref{II0r}), (\ref{II1r}).

\begin{table}[!ht]
\begin{center}
\begin{tabular}{|c|r|c|}
\hline
$P$ & $\hfill W \hfill$ & $Pg_W$ \\
\hline
2& $AB$ & $\hfill $1$ \hfill$ \\
\hline
3& $AAB$ & $1-\mu$ \\
 & $ABB$ & $1-\mu^3$ \\
\hline
4& $AAAB$ & $1-\mu$ \\
 & $AABB$ & $1-\mu^2$ \\
 & $ABBB$ & $1+\mu$ \\
\hline
5& $AAAAB$ & $1-\mu$ \\
 & $AAABB$ & $1-\mu^2$ \\
 & $AABAB$ & $(1-\mu)(2+\mu)$ \\
 & $AABBB$ & $(1-\mu)(1+\mu)^2$ \\
 & $ABABB$ & $(1-\mu^2)(2+\mu^2)$ \\
 & $ABBBB$ & $(1-\mu^5)(1+\mu)$ \\
\hline
6& $AAAAAB$ & $1-\mu$ \\
 & $AAAABB$ & $1-\mu^2$ \\
 & $AAABAB$ & $(1-\mu)(2+\mu)$ \\
 & $AAABBB$ & $(1-\mu)(1+\mu)^2$ \\
 & $^\star AABABB$ & $2(1-\mu^2)$ \\
 & $^\star AABBAB$ & $(1-\mu)(2+\mu+\mu^2)$ \\
 & $AABBBB$ & $(1-\mu^3)(1+\mu)$ \\
 & $ABABBB$ & $2+\mu$ \\
 & $ABBBBB$ & $1+\mu+\mu^2$ \\
\hline
\end{tabular}
\caption{
Exact expressions of $P$ times the gain amplitude $g_W$ for
all periodic history-dependent Parrondo games $W$
with primitive period $P\le6$.}
\label{IIpg}
\end{center}
\end{table}

The following characteristics emerge from the results listed in Table~\ref{IIpg}.
For all periodic games,
the product $Pg_W$
is a polynomial in $\mu$ with integer coefficients.
At variance with the case of capital-dependent games,
the gain amplitude is not invariant under time reversal.
The two unit cells of period 6 marked by asterisks
are the shortest ones exhibiting this lack of symmetry.
They are time-reversed of each other and have different amplitudes.

The situation where $\mu=0$, i.e., $a=1/2$, is very special.
Indeed, for $u=v=0$ both Rule~$A$ and Rule~$B$ correspond to symmetric random walks.
This is the only case where an exact expression of the gain amplitude $g_W$
can be obtained for all periodic games, namely
\beq
g_W=\frac{\nu}{P},
\label{nuP}
\eeq
where $\nu$ is the number of blocks of letters $A$
(or, equivalently, of blocks of letters $B$) in the unit cell $W$.
In other words,~$2\nu$ is the number of interfaces between blocks per period.

The maximal gain amplitude is reached for either the first or the second
of the periodic games listed in Table~\ref{IIpg},
according to values of $a$,
namely
\beq
g^\max=\left\{\matrix{
g_{AB}=\frad{1}{2}\hfill\quad & \hbox{for }0<a<3/4,\cr
g_{AAB}=\frad{2a}{3}\quad & \hbox{for }3/4<a<1.
\label{IIgmax}
}\right.
\eeq
It has been checked by means of an exhaustive enumeration
that no higher gain is reached for periods up to $P=30$.
For $a=1/2$, the above result is a consequence of~(\ref{nuP}),
as the ratio $\nu/P$ reaches its maximum $1/2$
for the periodic game $AB$.
It however comes as a surprise that~$AB$ remains the optimal game over three quarters
of the range of the parameter $a$.

We again make a digression out of the weak-contrast regime
in order to look at the maximal gain of the history-dependent Parrondo game
all over its parameter space.
For fixed~$a$,
the periodic games $AB$ and $AAB$ reach their respective highest gain,
namely
\beq
G^\max_{AB}=\frac{(1-a)^2}{2(1+a)^2},\quad
G^\max_{AAB}=\frac{(1-a)^2}{6},
\label{his}
\eeq
at both midpoints D and~G (see Figure~\ref{IIefgh}).
For fixed $a$ in the range $a>1/2$,
the maximal gain -- over~$b$ and $c$ and over all possible rule patterns --
is always the larger of both expressions given in~(\ref{his}).
The situation is however different for $a<1/2$.
There, the optimal periodic game undergoes an infinite sequence of transitions
towards longer and longer periods as $a$ becomes smaller and smaller.
The absolute maximal gain is given by
\beq
G^\max\to1.
\eeq
This limiting value was already encountered in the framework of random games (see~(\ref{avetriv})).
It is approached in the coupled singular limit where $a\to0$,
whereas the periods of optimal rule patterns diverge.

\subsection{Aperiodic games}
\label{IIwape}

The formalism of Section~\ref{IIwgen} extends to any aperiodic game,
either deterministic or random.
We do not have any analytical result such as (\ref{gaperes}).
Nevertheless, the recursions (\ref{II0r}), (\ref{II1r})
can be iterated by numerical means for any given aperiodic sequence.
Because of the exponential damping property of these recursions,
very accurate numerical values of the amplitude $g$ can be obtained,
especially in situations where the fluctuations $\delta_n$ defined in~(\ref{flucs}) are small.

\begin{figure}[!ht]
\begin{center}
\includegraphics[angle=0,width=.8\linewidth]{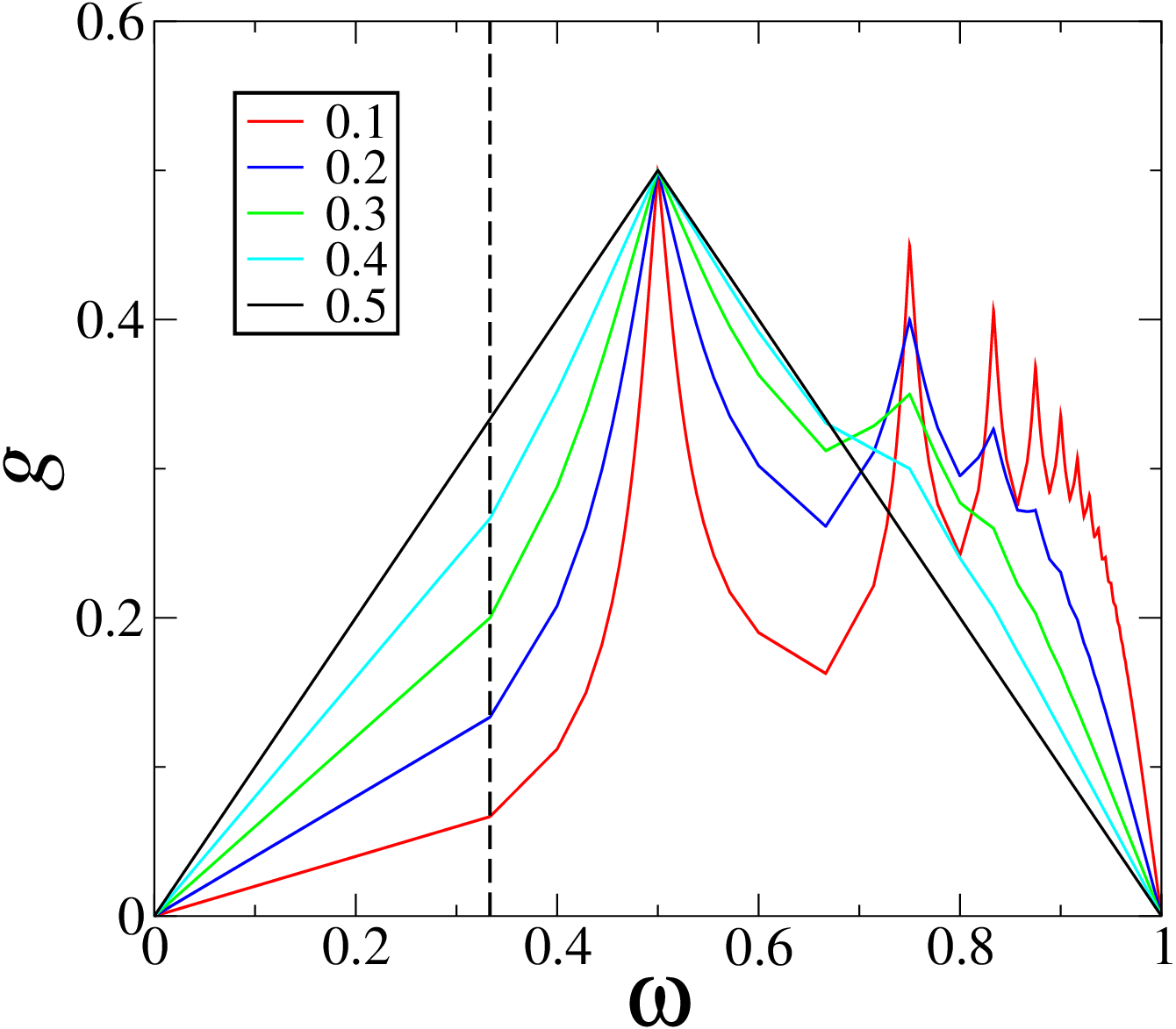}
\vskip 10pt
\includegraphics[angle=0,width=.8\linewidth]{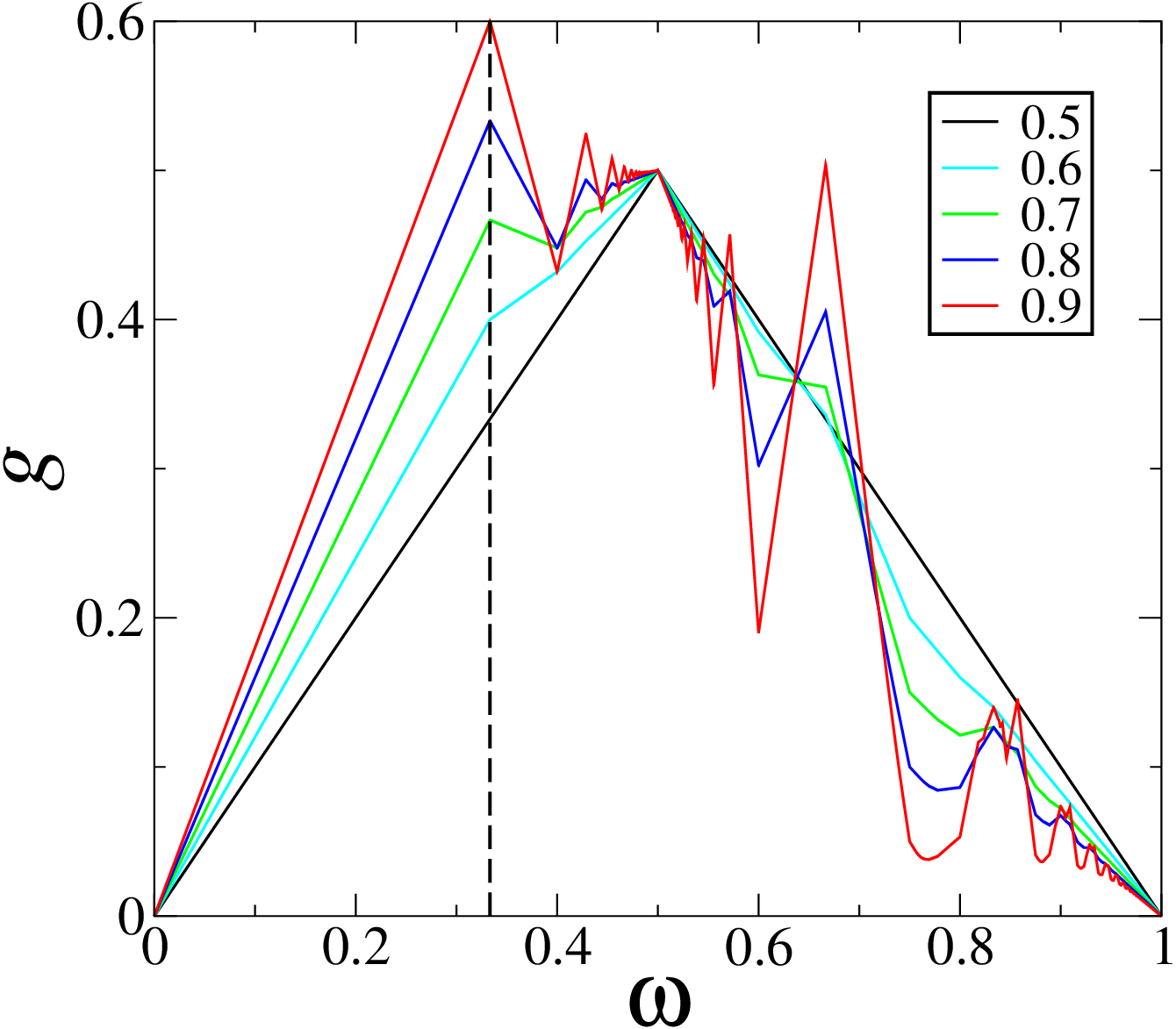}
\vskip 8pt
\caption{
Dependence of the gain amplitude $g$
of the history-dependent Parrondo game
on the rotation number $\o$ defining the cut-and-project sequence,
for several values of $a$ (see legend).
Upper panel: $a\le1/2$.
Lower panel: $a\ge1/2$.
Vertical dashed lines: upper edge ($\o=1/3$) of validity of the linear law~(\ref{glinglin}).}
\label{IIcut}
\end{center}
\end{figure}

We again consider the cut-and-project aperiodic game introduced in Section~\ref{caps}.
Figure~\ref{IIcut} shows plots of the gain amplitude $g$
against the rotation number $\o$ defining the cut-and-project sequence,
for several values of the parameter $a$.
Curves for $a\le1/2$ and $a\ge1/2$
are shown in two separate panels, for the sake of clarity.

For $a=1/2$, the result (\ref{nuP}) translates to
\beq
g=\left\{\matrix{
\o\hfill & \hbox{for }0\le\o\le1/2,\cr
1-\o\quad & \hbox{for }1/2\le\o\le1.
}\right.
\eeq
The corresponding triangular shape is shown in black in both panels of Figure~\ref{IIcut}.
For $\o\le1/3$,
all letters~$B$ are isolated and separated from each other by at least two letters $A$.
Setting $k=0$ in the expression (\ref{IIab4}),
we predict that each letter $B$ in the sequence brings a contribution $1-\mu=2a$ to the gain.
We thus obtain the linear law
\beq
g=2a\o\quad(0\le\o\le1/3),
\label{glinglin}
\eeq
that is clearly visible to the left of the vertical dashed lines in both panels of Figure~\ref{IIcut}.
As a general rule, the dependence of the amplitude $g$ on the rotation number $\o$
exhibits more and more pronounced fine details as $\abs{\mu}$ grows,
i.e., as $a$ departs from $1/2$ on both sides.
Red curves correspond to the largest values of~$\abs{\mu}$,
namely $\mu=4/5$ ($a=1/10)$ in the upper panel,
and $\mu=-4/5$ ($a=9/10)$ in the lower panel.

\begin{figure}[!ht]
\begin{center}
\includegraphics[angle=0,width=.8\linewidth]{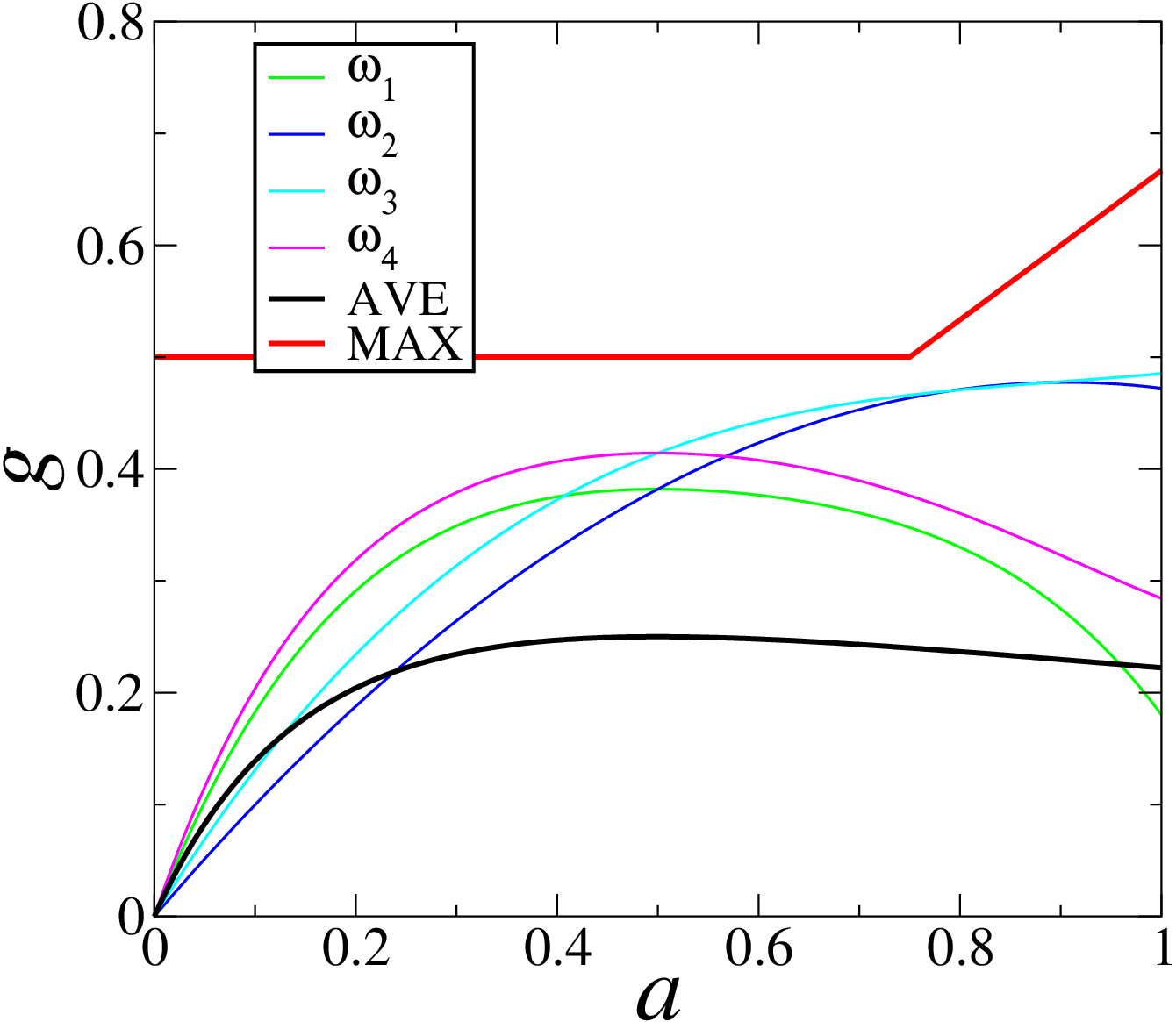}
\vskip 6pt
\caption{
Dependence of the gain amplitude $g$
of the history-dependent cut-and-project game
on the parameter $a$,
for four typical irrational rotation numbers
(see legend).
AVE: amplitude $\ave{g}$ of the uniformly random game (see~(\ref{IIgavg})).
MAX: maximal amplitude~$g^\max$ (see~(\ref{IIgmax})).}
\label{IIvar}
\end{center}
\end{figure}

Figure~\ref{IIvar}
shows the dependence of the amplitude~$g$ on the parameter $a$
for four typical irrational rotation numbers:
$\o_1=1/\tau=(\sqrt{5}-1)/2$,
$\o_2=1/\tau^2=(3-\sqrt{5})/2$,
$\o_3=\sqrt{2}-1$,
$\o_4=2-\sqrt{2}$.
The first two numbers are related to Fibonacci (or golden-mean) sequences,
the last two to octonacci (or silver-mean) sequences
(see~\cite{janot,sene} for overviews).
The amplitude $\ave{g}$ of the uniformly random game (see~(\ref{IIgavg}))
and the maximal amplitude $g^\max$ (see~(\ref{IIgmax}))
are also shown for comparison.

\section{Overview}
\label{disc}

This paper is aimed at being part of a special issue on the theory of disordered systems.
It has been written in a fully self-contained manner.
Of course, we have no claim to compete with either historical~\cite{R1,R2,R3,R4}
or very recent~\cite{R5} reviews on Parrondo games and Parrondo's paradox.
Our motivation was to draw on the analogy between
the temporal products of non-commuting Markov matrices involved in the study of Parrondo games
and the spatial products of non-commuting transfer matrices
which are ubiquitous in the physics of 1D disordered systems.
There are many similarities as well as differences between both situations.
The most salient common feature is that the non-commutativity of the matrix products
ascribes a crucial role to the order of factors,
representing either the rule pattern in Parrondo games
or the positions of impurities in disordered chains.
Markov matrices however enjoy a very specific property.
They conserve probability, and so the entries of products of Markov matrices are bounded by unity.
The concept of Lyapunov exponent,
which is otherwise central in most situations involving products of random matrices,
is therefore virtually useless in the present setting.

The investigations of Parrondo games
reported here have been freely inspired by the theory of 1D disordered systems.
We have dealt with both capital-dependent and history-dependent
Parrondo games on the same footing in a systematic way,
by means of a mapping onto a random walker on the 1D lattice.
Within this unifying framework,
the gain $G$ of the player identifies with the velocity
of the walker's ballistic motion.
For definiteness,
we have chosen one paradigmatic game in each class,
and focussed our attention onto the neutral situation where each rule, when played alone, is fair
($G_A=G_B=0$).
The main emphasis is on the dependence of the gain on the remaining free parameters and,
more importantly, on the game, i.e., the rule pattern,
be it periodic or aperiodic, deterministic or random.

One of the most original sides of this work is the identification of weak-contrast regimes
for both classes of Parrondo games considered here,
and a detailed quantitative investigation of the gain in the latter scaling regimes.
For the capital-dependent game mod 3 introduced in Section~\ref{I},
encompassing Parrondo's historical example,
one single asymmetry parameter $v$ characterizes the neutral situation.
The weak-contrast regime, studied in Section~\ref{Iweak},
corresponds to $v\to0$, where the gain of a generic game
scales as $G\approx gv^3$.
For the two-step history-dependent game introduced in Section~\ref{II},
the neutral situation is richer, as it depends on three parameters.
The weak-cont\-rast regime, studied in Section~\ref{IIweak},
corresponds to both relevant asymmetry parameters $u$ and $v$ being simultaneously small.
The gain of a generic game now scales as $G\approx guv$.
For both classes of games, the determination of the gain amplitude~$g$
has been reduced to the solution of two coupled linear recursions.
This reduction allowed us to derive a wealth of novel results on both classes of Parrondo games.
It is expected that more complex Parrondo games,
with either $K>3$ for capital-dependent games or $Q>2$ for history-dependent games,
admit weak-contrast scaling regimes in full generality,
even though the number of remaining relevant parameters in those regimes
grows very fast with the complexity of the game.

\bibliography{paper.bib}

\end{document}